\documentclass[utf8]{frontiersSCNS} % for Science, Engineering and Humanities and Social Sciences articles
\usepackage{url,hyperref,lineno,microtype,subcaption,soul}
\usepackage[onehalfspacing]{setspace}
\usepackage{color}
\newcommand{\red}{\textcolor{black}}

\newcommand{\be}{\begin{equation}} 
\newcommand{\ee}{\end{equation}}
\newcommand{\bea}{\begin{eqnarray}} 
\newcommand{\eea}{\end{eqnarray}}
\def\keyFont{\fontsize{8}{11}\helveticabold}
\def\firstAuthorLast{Bi {et~al.}} %use et al only if is more than 1 author
\def\Authors{Hongjie Bi\,$^{1,2}$, Matteo di Volo\,$^{1}$ and Alessandro Torcini\,$^{1,3,*}$}
\def\Address{$^{1}$ CY Cergy Paris Universit\'e, Laboratoire de Physique Th\'eorique et Mod\'elisation, CNRS, UMR 8089, 95302 Cergy-Pontoise cedex, France \\
$^{2}$ Okinawa Institute of Science and Technology, Onna-son, Okinawa 904-0495, Japan \\
$^{3}$ CNR - Consiglio Nazionale delle Ricerche - Istituto dei Sistemi Complessi, via Madonna del Piano 10, 50019 Sesto 
Fiorentino, Italy }
\def\corrAuthor{Alessandro Torcini}
\def\corrEmail{alessandro.torcini@cyu.fr}
   
\begin{document}
\onecolumn
\firstpage{1}

\title[Asynchronous and coherent dynamics in balanced \red{spiking} networks]{Asynchronous and coherent dynamics in balanced excitatory-inhibitory \red{spiking networks} }

\author[\firstAuthorLast ]{\Authors} %This field will be automatically populated
\address{} %This field will be automatically populated
\correspondance{} %This field will be automatically populated

\extraAuth{}% If there are more than 1 corresponding author, comment this line and uncomment the next one.
%\extraAuth{corresponding Author2 \\ Laboratory X2, Institute X2, Department X2, Organization X2, Street X2, City X2 , State XX2 (only USA, Canada and Australia), Zip Code2, X2 Country X2, email2@uni2.edu}

\maketitle

%=====================================================================================
\begin{abstract}

Dynamic excitatory-inhibitory (E-I) balance is a paradigmatic mechanism invoked to
explain the irregular low firing activity observed in the cortex. 
However, we will show that the E-I balance can be at the origin of other regimes observable in the brain.
The analysis is performed by combining extensive simulations of sparse
E-I networks composed of $N$ spiking neurons with analytical 
investigations of low dimensional neural mass models. 
The bifurcation diagrams, derived for the neural mass model, allow to 
classify the possible asynchronous and coherent behaviours emerging in balanced E-I 
networks with structural heterogeneity for any finite in-degree $K$.
\red{Analytic mean-field results show that both supra and sub-threshold balanced asynchronous regimes
are observable in our system in the limit $ N >> K >> 1$.}
Due to the heterogeneity the asynchronous states are characterized \red{at the microscopic level}
by the splitting of the neurons in three groups: silent, fluctuation and mean driven. These features are consistent 
with experimental observations reported for heterogeneous neural circuits.
The coherent rhythms observed in our system can range from periodic and quasi-periodic collective
oscillations (COs) to coherent chaos. These rhythms are characterized by regular or irregular temporal fluctuations 
joined to spatial coherence somehow similar to 
coherent fluctuations observed in the cortex over multiple spatial scales.
The COs can emerge due to two different mechanisms. A first mechanism \red{analogous} to the
pyramidal-interneuron gamma (PING) one, usually invoked for the emergence of $\gamma$-oscillations.
The second mechanism is intimately related to the presence of current fluctuations, which
sustain COs characterized by an essentially simultaneous bursting of the two populations. 
We observe period-doubling cascades involving the PING-like COs finally leading to the
appearance of coherent chaos. Fluctuation driven COs are usually observable in our system as quasi-periodic collective motions
characterized by two incommensurate frequencies. \red{However, for sufficiently strong current fluctuations
these collective rhythms can lock. This represents a novel mechanism of frequency locking in neural populations
promoted by intrinsic fluctuations. COs are observable for any finite in-degree $K$, however their existence in the limit 
$N >> K >> 1$ appears as uncertain. }

\tiny
\keyFont{\section{Keywords:} Balanced spiking neural populations, sparse excitatory-inhibitory network, asynchronous dynamics, collective oscillations, neural mass models, quadratic integrate-and-fire neurons, structural heterogeneity, coherent chaos} 
 %All article types: you may provide up to 8 keywords; at least 5 are mandatory.
\end{abstract}
%=====================================================================================

\section{Introduction}

Cortical neurons are subject to a continuous bombardment from thousands of pre-synaptic neurons, mostly 
pyramidal ones, evoking post-synaptic potentials of sub-millivolt or millivolt amplitudes \citep{destexhe1999,bruno2006,lefort2009}.
This stimulation would induce an almost constant depolarization of the neurons leading to a regular firing,
However, cortical neurons fire quite irregularly and with low firing rates \citep{softky1993}.
This apparent paradox can be solved by introducing the concept of a balanced network, 
where excitatory and inhibitory synaptic currents are approximately balanced and the neurons are kept near their firing threshold crossing it at random times \citep{shadlen1994,shadlen1998}. 
However, the balance should naturally emerge in the network without fine tuning of the parameters
and the highly irregular firing observed {\it in vivo} should be maintained also for large number
of connections (in-degree) $K >> 1$. This is possible by considering a sparse excitatory-inhibitory (E-I) neural network composed
by $N$ neurons and characterized by an average in-degree $K << N$ and by synaptic couplings
scaling as $1/\sqrt{K}$ \citep{bal1}. This scaling as well as many other key predictions of the theory
developed in \citep{bal1} have been recently confirmed by experimental measurements {\it in vitro}
on a neural culture optogenetically stimulated \citep{barral2016}. Furthermore, the authors in
\citep{barral2016} have shown that the major predictions of the seminal theory \citep{bal1} 
hold also under conditions far from the asymptotic limits where $K$ and $N$ are large.

The dynamics usually observable in balanced neural networks  is asynchronous and characterized
by irregular neural firing joined to stationary firing rates \citep{bal1,bal2,bal3,wolf,ullner2020}.
However, asynchronous regimes characterized by mean driven sub-Poissonian statistics as well as by 
super-Poissonian one have been reported in balanced homogeneous and heterogeneous networks \citep{lerchner2006, ullner2020}.
Furthermore, regular and irregular collective oscillations (COs) have been shown to emerge in balanced
networks composed of rate models \citep{bal1} as well as of spiking neurons  \citep{brunel2000,ostojic,ullner2018,matteo,bi2020}.
The balanced asynchronous irregular state has been experimentally observed both {\it in vivo} and {\it in vitro} \citep{shu2003,haider2006} and dynamic balance of excitation and inhibition is observable  in the neocortex
across all states of the wake-sleep cycle, in both human and monkey \citep{dehghani2016dynamic}. 
\red{However, this is not the unique balanced state observable in neural systems. In particular,
balancing of excitation and inhibition appears to be crucial for the emergence of cortical oscillations 
\citep{okun2008,isaacson2011,le2016} as well as for the instantaneous modulation of 
gamma oscillations frequency in the hippocampus \citep{atallah2009}. Moreover, 
balancing of excitation and inhibition is essential for the generation of respiratory rhythms in the brainstem
\citep{ramirez2018} and for the rhythmic activity of irregular firing motoneurons in the spinal cord 
of the turtle  \citep{berg2007,berg2019}.
}
 
In this work we characterize in details the asynchronous regimes and the emergence of
COs (population rhythms) in  E-I balanced networks with structural heterogeneity. In particular, we consider sparse random
networks of quadratic integrate-and-fire (QIF) neurons \citep{ermentrout1986}
pulse coupled via instantaneous post-synaptic potentials.
We compare numerical findings with analytical results obtained in the mean-field (MF) limit
by employing an effective low-dimensional neural mass model recently developed
for sparse QIF networks \citep{montbrio2015, matteo, bi2020}.

In the asynchronous regime, our analytical MF predictions are able reproduce the mean membrane potentials
and the population firing rates of the structurally heterogeneous network for any finite $K$ value. Furthermore, 
in the limit $N >> K >> 1$ we analytically derive the asymptotic MF values of the population firing rates
as well as of the effective input currents. This analysis shows that the system always achieve
a balanced dynamics, whose supra or sub-threshold nature is determined by the model parameters.
Detailed numerical investigations of the microscopic dynamics allow to identify three different groups of neurons,
whose activity is essentially controlled by their in-degrees and by the effective input currents.
 
In the balanced network we have identified three types of COs depending on the corresponding 
solution displayed by the neural mass model. The first type, termed O$_P$ emerges 
in the MF via a Hopf bifurcation from a stable focus solution. 
These COs gives rise to collective chaos via a period-doubling sequence of bifurcations.
Another type of CO, already reported for purely inhibitory networks
\citep{matteo}, denoted as O$_F$ corresponds in the MF to a stable focus characterized 
by relaxation oscillations towards the fixed point that in the sparse network become noise
sustained oscillations due to fluctuations in the input currents.
The last type of COs identified in the finite network are named O$_S$ and characterized by an abnormally synchronized
dynamics among the neurons, the high level of synchronization prevents their representation in the MF formulation \citep{montbrio2015}.

O$_P$ and O$_S$ emerge as sustained oscillations in the network thanks to
a mechanism similar to that reported for pyramidal-interneuron gamma (PING) rhythms \citep{whittington2011}
despite the frequency of these oscillations are not restricted to the $\gamma$ band. Excitatory neurons
start to fire followed by the inhibitory ones and the peak of activity of the excitatory population
precedes that of the inhibitory one of a time delay $\Delta t$. Furthermore, $\Delta t$ tends to
vanish when the amplitude of the current fluctuations in the network increases. Indeed, for O$_F$
oscillations, which cannot emerge in absence of current fluctuations, no delay has been observed between
the activation of excitatory and inhibitory population. \red{A last important question that we tried to address 
in our work is if the COs, observable for any finite $K$, are still present in the limit $N >> K >> 1$. 
}

The paper is organized as follows. Section 2 is devoted to the introduction of the network model
and of the corresponding effective neural mass model, as well as of the microscopic and 
macroscopic indicators employed to characterize the neural dynamics.  In the same Section,
the stationary solutions for the balanced neural mass model are analytically obtained as finite in-degree
expansion and their range of stability determined. \red{The macroscopic dynamical regimes 
emerging in our network are analysed in Section 3. In particular, we report bifurcation phase diagrams 
obtained from the neural mass model displaying the possible dynamical states as well as network simulations.
Particular attention is devoted in this Section to the analysis of the
asynchronous balanced state for structurally heterogeneous networks
and to the emergence of the different types of COs observable at finite in-degrees.
A discussion of the obtained results and conclusions
are reported in Section 4.}

 %=====================================================================================
 \section{Models and Dynamical Indicators}
\vspace{0.2cm}
\subsection{Network Model}
\vspace{0.2cm}
We consider two sparsely coupled excitatory and inhibitory populations composed of $N^{(e)}$ and $N^{(i)}$ 
quadratic integrate-and-fire (QIF) neurons, respectively.~\citep{ermentrout1986}.
The evolution equation for the membrane potentials $v_j^{(e)}$ and $v_j^{(i)}$
of the excitatory and inhibitory neurons can be written as:
\begin{subequations} \label{eq:1}
\bea
\tau_{m} \dot{v}_{j}^{(e)}  &=&  \left(v_{j}^{(e)}\right)^2+I^{(e)} 
+ 2\tau_m \left[ g^{(ee)} \sum_{l | t_l^{(n)} < t } \epsilon_{jl}^{(ee)} \delta(t-t_l^{(n)}) 
- g^{(ei)} \sum_{k | t_k^{(m)} < t} \epsilon_{jk}^{(ei)} \delta(t-t_k^{(m)})
\right]
\\
\tau_{m} \dot{v}_{j}^{(i)}  &=&  \left(v_{j}^{(i)}\right)^2+I^{(i)} 
+ 2\tau_m \left[ g^{(ie)} \sum_{l | t_l^{(n)} < t } \epsilon_{jl}^{(ie)} \delta(t-t_l^{(n)}) 
- g^{(ii)} \sum_{k | t_k^{(m)} < t} \epsilon_{jk}^{(ii)} \delta(t-t_k^{(m)})
\right]  
\eea
\end{subequations}
where $\tau_{m}=20$ ms is the membrane time constant that we set identical for excitatory and inhibitory neurons, $I^{(e)}$ ($I^{(i)}$) is the external DC current acting on excitatory (inhibitory) population, $g^{(\alpha\beta)}$ represents the synaptic coupling strengths between post-synaptic neurons in population $\alpha$ and pre-synaptic ones in population $\beta$, with $\alpha,\beta \in \{e,i\}$. The elements of the adjacency matrices $\epsilon_{jk}^{(\alpha\beta)}$  are equal to $1$  (0) if a connection from a pre-synaptic neuron $k$ of population $\beta$ towards a post-synaptic neuron $j$ of population $\alpha$, exists (or not). Furthermore,
$k_j^{(\alpha \beta)} =\sum_k\epsilon_{jk}^{(\alpha\beta)}$ is the number of pre-synaptic neurons in population $\beta$ connected to neuron $j$ in population $\alpha$, or in other terms its in-degree restricted to population $\beta$. The emission of the $n$-th spike emitted by neuron $l$ of population $\alpha$ occurs at time $t_l^{(n)}$ whenever the membrane potential ${v}_{l}^{(\alpha)} ({t_l^{(n)}}^-) \to \infty$, while the reset mechanism is modeled by setting ${v}_{l}^{(\alpha)} ({t_l^{(n)}}^+) \to -\infty$ immediately after the spike emission. The post-synaptic potentials are assumed to be $\delta$-pulses and the synaptic transmissions to be instantaneous. 
The equations \eqref{eq:1} can be formally rewritten as 
\begin{equation} 
\label{eq_ieff}
\tau_{m} \dot{v}_{j}^{(e)}  =  \left(v_{j}^{(e)}\right)^2+ i^{(e)}_{eff,j} \qquad , \qquad 
\tau_{m} \dot{v}_{j}^{(i)}  =  \left(v_{j}^{(i)}\right)^2 + i^{(i)}_{eff,j} \qquad; 
\end{equation}
where $i^{(e)}_{eff,j}$ ($i^{(i)}_{eff,j}$) represents the instantaneous excitatory (inhibitory) effective currents,
which includes the external DC current as well as the synaptic currents due to the recurrent connections.

We consider the neurons within excitatory and inhibitory population as randomly connected, with in-degrees $k^{(\alpha\alpha)}$ distributed according to a Lorentzian distribution 
\begin{equation}\label{eq:2}
P(k^{(\alpha \alpha)})= \frac{\Delta^{(\alpha \alpha)}_k}{(k^{(\alpha\alpha)}-K^{(\alpha \alpha)})^2+{\Delta^{(\alpha\alpha)}_k}^2}
\end{equation}
peaked at $K^{(\alpha \alpha)}$ and with a half-width half-maximum (HWHM) $\Delta^{(\alpha \alpha)}_k$, this latter parameter measures the level of structural heterogeneity in each population. For simplicity, we set $K^{(ee)}=K^{(ii)} \equiv K$. Furthermore, we assume that also neurons from a population $\alpha$ are randomly connected
to neurons of a different population $\beta \ne \alpha$. However, in this case we consider no structural heterogeneity with in-degrees fixed to a constant value $K^{(ei)} = K^{(ie)} = K$.
We have verified that by considering Erd\"os-Renyi distributed in-degrees $K^{(ei)}$ and $K^{(ie)}$
with average $K$ does not modify the observed dynamical behaviour. 
 
The DC current and the synaptic coupling are rescaled with the median in degree as $I^{(\alpha)}=\sqrt{K}I_0^{(\alpha)}$ and $g^{(\alpha\beta)}=g_0^{(\alpha\beta)}/\sqrt{K}$, as done in previous works to obtain a self-sustained balanced dynamics for $N >> K >> 1$ \citep{bal1,bal2,bal3,bal4}. The structural heterogeneity parameters
are rescaled as $\Delta^{(\alpha \alpha)}_k=\Delta^{(\alpha \alpha)}_0\sqrt{K}$ in analogy to Erd\"os-Renyi networks. The choice of the Lorentzian distribution for the $k^{(\alpha \alpha)}$ is needed in order to obtain an effective MF description for the microscopic dynamics \citep{matteo,bi2020} as detailed in the next section. 
%However this choice does not modify the global dynamics and similar results
%can be obtained by considering Gaussian and Erd\"os-Reniy distributions \citep{bi2020}.

The microscopic activity can be analyzed by considering the inter-spike interval (ISI) distribution as characterized by the coefficient of variation $cv_i$ for each neuron $i$,
which is the ratio between the standard deviation and the mean of the ISIs associated to the train of spikes emitted by the considered neuron. To characterize the macroscopic dynamics of each population $\alpha$ we measure the average coefficient of variation 
$CV^{(\alpha)} = \sum_{i=1}^{N^{(\alpha)}} cv_i/N^{(\alpha)}$, the mean membrane potential $V^{(\alpha)}(t) = \sum_{i=1}^{N^{(\alpha)}} v^{(\alpha)}_i(t)/N^{(\alpha)}$ and the population firing rate $R^{(\alpha)}(t)$, corresponding to the number of spikes emitted within population $\alpha$ per unit of time and per neuron.

Furthermore, the level of coherence in the neural activity of population $\alpha$ can be quantified in terms of the following indicator \citep{golomb}
\begin{eqnarray}\label{eq:3}
\rho^{(\alpha)} = \left(\frac{\sigma_{V^{(\alpha)}}^{2}}{\sum_{i=1}^{N^{(\alpha)}}\sigma_{i}^{2}/N^{(\alpha)}}\right)^{1/2}
\end{eqnarray}
where $\sigma_{V^{(\alpha)}}$ is the standard deviation of the mean membrane potential, $\sigma_{i}^{2}=\left< (v^{(\alpha)}_{i})^{2}\right>-\left< v^{(\alpha)}_{i}\right>^{2}$ and $\left< \cdotp \right>$ denotes a time average. A perfect synchrony corresponds to 
$\rho^{(\alpha)} = 1$, while an asynchronous dynamics to a vanishing small $\rho^{(\alpha)} \approx O(1/\sqrt{N^{(\alpha)}})$.

The frequencies associated to collective motions can be identified by measuring the power spectra $S(\nu)$ of the mean membrane potentials $V(t)$ of the whole network. In case of a periodic motion the position of the main peak $\nu_{CO}$ represents the frequency of the COs, while for quasi-periodic motions the spectrum is characterized by many peaks that can be obtained as a linear combination of two fundamental frequencies $(\nu_1,\nu_2)$. The spectra obtained in the present case always exhibits also a continuous background due
to the intrinsic fluctuations present in the balanced network. The power spectra have been obtained by calculating the temporal Fourier transform of $V(t)$ sampled at time intervals of 10 ms. Time traces composed of 10000 consecutive intervals have been considered to estimate the spectra, which are obtained at a frequency resolution of $\Delta \nu = 0.01$ Hz. Finally, the power spectra have been averaged over five independent realizations of the random network.

The network dynamics is integrated by employing an Euler scheme with time step $dt = 0.0001$ ms, while time averages and fluctuations are usually estimated on time intervals $T_s \simeq 100$ s, after discarding transients $T_t \simeq 10$ s. Usually we consider networks composed of $N^{(e)}=10000$ excitatory and $N^{(i)}=2500$ inhibitory neurons.

%=====================================================================================
\vspace{0.5cm}
\subsection{Effective neural mass model}
\vspace{0.5cm}

\red{In this sub-section we derive a low dimensional effective neural mass formulation for the spiking network \eqref{eq:1}
by following \citep{montbrio2015}. In such article the authors obtained an exact mean-field model for a globally coupled heterogeneous population of QIF neurons by generalizing to neural systems a reduction methodology previously developed for phase-coupled oscillators by Ott and Antonsen \citep{ott2008}. In particular, the neural mass model can be obtained by performing a rigorous mathematical derivation
from the original spiking network in the limit $N \to \infty$ by assuming that the heterogeneity present in the network, which can be
either neuronal excitabilities or synaptic couplings, are distributed as Lorentzians.  This mean-field reduction methodology
gives rise to a neural mass model written in terms of of only two collective variables: the mean membrane potential $V$ and the instantaneous population rate $R$. For sufficiently large network size, the agreement between the simulation results
and the neural mass model is impressive as shown in \citep{montbrio2015}  and in several successive publications.
} 

\red{The detailed derivation of the neural mass models from the corresponding spiking networks can be found in \citep{montbrio2015},
here we limit to report its expression for a fully coupled homogeneous network of QIF neurons with synaptic couplings randomly distributed according to a Lorentzian}: 
\begin{subequations}\label{mf_0}
\bea
\tau_m \dot{R} &=&  2 R V + \frac{\Gamma}{\pi} R \\
\tau_m \dot{V} &=& {V^2 + I} + {\bar g}   \tau_m R -(\pi \tau_m R)^2
\eea
\end{subequations}
where ${\bar g}$ is the median and $\Gamma$ the HWHM of the Lorentzian distribution of the synaptic couplings.

Such formulation can be applied to the random sparse network studied in this paper, indeed as shown in \citep{matteo, bi2020} for a single sparse inhibitory population the quenched disorder associated to the in-degree distribution can be rephrased in terms of random synaptic couplings.  Namely, each neuron $i$ in population $\alpha$ is subject to currents of amplitude $g_0^{(\alpha \beta)} k_i^{(\alpha \beta)} R^{(\beta)}/(\sqrt{K})$ proportional to their in-degrees $k_i^{(\alpha \beta)}$, with $\beta \in \{e,i\}$. Therefore we can consider the neurons as fully coupled, but with random values of the couplings distributed as Lorentzian  of median $g^{(\alpha \beta)}_0 \sqrt{K}$  and HWHM $g^{(\alpha \beta)}_0 \Delta^{(\alpha \beta)}_0$.  
 
The neural mass model corresponding to the spiking network  \eqref{eq:1} can be written as follows:
\begin{subequations}\label{eq:5a}
\bea
\tau_m\dot{R}^{(e)} &=& R^{(e)}\left[2V^{(e)}+g_{0}^{(ee)}\frac{\Delta_{0}^{(ee)}}{\pi}\right]  \label{rate}\\  
\tau_m\dot{V}^{(e)} &=& \left[V^{(e)}\right]^2 -\left[\pi R^{(e)}\tau_{m}\right]^2
+ \sqrt{K}\left[I_0^{(e)} +(g_{0}^{(ee)}R^{(e)}-g_{0}^{(ei)}R^{(i)})\tau_m\right] \\
\tau_m\dot{R}^{(i)} &=& R^{(i)}\left[2V^{(i)}+g_{0}^{(ii)}\frac{\Delta_{0}^{(ii)}}{\pi}\right]  \label{rate1}\\  
\tau_m\dot{V}^{(i)} &=& \left[V^{(i)}\right]^2 -\left[\pi R^{(i)}\tau_{m}\right]^2  + \sqrt{K}\left[I_0^{(i)} +(g_{0}^{(ie)}R^{(e)}-g_{0}^{(ii)}R^{(i)})\tau_m\right] \quad ;
\eea
\end{subequations}
where we have set $\Delta^{(e i)}_0 = \Delta^{(i e)}_0 =0$, since we have
assumed that the connections among neurons of different populations are random but with 
a fixed in-degree $K^{(ei)} = K^{(ie)} = K$.

\vspace{0.5cm}
\subsubsection{Stationary Solutions}
\vspace{0.5cm}

The stationary solutions $\{\overline{V}^{(e)}, \overline{V}^{(i)}, \overline{R}^{(e)}, \overline{R}^{(i)}\}$  of \eqref{eq:5a} can be explicitly obtained for the mean membrane potentials as 
\begin{equation}
\overline{V}^{(e)}=-\frac{g_0^{(ee)} \Delta_{0}^{(ee)}}{2 \pi} \qquad, \qquad
\overline{V}^{(i)}=-\frac{g_0^{(ii)} \Delta_{0}^{(ii)}}{2 \pi} \qquad ;
\label{stat_V}
\end{equation}
while the instantaneous population rates are the solutions of the following quadratic system
\begin{subequations}\label{stat_R}
\bea
g_{0}^{(ee)} \overline{R}^{(e)} \tau_m-g_{0}^{(ei)} \overline{R}^{(i)}\tau_m &=& - I_0^{(e)} + 
\varepsilon \left\{\left[ \pi \overline{R}^{(e)}\tau_{m}\right]^2 - \left[ \overline{V}^{(e)} \right]^2 \right\}
\\
g_{0}^{(ie)} \overline{R}^{(e)} \tau_m -g_{0}^{(ii)} \overline{R}^{(i)}\tau_m &=& - I_0^{(i)} + 
\varepsilon \left\{ \left[ \pi \overline{R}^{(i)}\tau_{m} \right]^2 -\left[\overline{V}^{(i)} \right]^2 \right\}
\eea
\end{subequations}
where $\varepsilon = 1/\sqrt{K}$ is a smallness parameter taking in account finite in-degree corrections. 
It is interesting to notice that the parameters controlling the structural heterogeneity $\Delta_0^{(ii)}$
and $\Delta_0^{(ee)}$ fix the stationary values of the mean membrane potentials reported in \eqref{stat_V}.
The solutions of \eqref{stat_R} can be exactly obtained and the associated bifurcations analysed by employing the software XPP AUTO developed for orbit continuation \citep{XPP2007}. 

For  sufficiently large $K$ one can obtain analytic approximations of the solution of \eqref{stat_R} by expanding the population rates as follows
\begin{equation}\label{exp}
\overline{R}^{(\alpha)} =  \overline{R}^{(\alpha)}_0 + \varepsilon \overline{R}^{(\alpha)}_1 + \varepsilon^2 \overline{R}^{(\alpha)}_2 + \varepsilon^3 \overline{R}^{(\alpha)}_3  + \dots  \qquad \alpha \in \{e,i\} \quad,
\end{equation}
by inserting these expressions in \eqref{stat_R}, and finally by solving order by order in $\varepsilon$.

The solutions at any order can be written as follows:
 
\begin{equation}
\label{balance}
\overline{R}^{(e)}_k \tau_m =\frac{N_k^{(e)}g_{0}^{(ii)}-N_k^{(i)}g_{0}^{(ei)}}{g_0^{(ei)}g_0^{(ie)}-g_0^{(ee)}g_0^{(ii)}} 
\quad , \quad
\overline{R}^{(i)}_k \tau_m = \frac{N_k^{(e)}g_{0}^{(ie)}-N_k^{(i)}g_{0}^{(ee)}}{g_0^{(ei)}g_0^{(ie)}-g_0^{(ee)}g_0^{(ii)}} \quad;
\end{equation}
where
\begin{subequations}\label{constant} 
\bea
N_0^{(\alpha)} &=& I_0^{(\alpha)} \enskip, \enskip N_1^{(\alpha)} = \left[ \overline{V}^{(\alpha)} \right]^2 -\left[ \pi \overline{R}^{(\alpha)}_0 \tau_{m}\right]^2 \\
N_{2j}^{(\alpha)} &=& -2\left[\pi \tau_{m} \right]^{2} \sum_{k=1}^j \left[\overline{R}^{(\alpha)}_{k-1}
\overline{R}^{(\alpha)}_{2j-k} \right] \\
N_{2j+1}^{(\alpha)} &=& -2\left[\pi \tau_{m} \right]^{2} \left\{ \left[ \sum_{k=1}^j \overline{R}^{(\alpha)}_{k-1}\overline{R}^{(\alpha)}_{2j+1-k}  \right]  + \frac{1}{2} \left[\overline{R}^{(\alpha)}_{j}\right]^{2}   \right\} 
\enskip {\rm for} \enskip j \ge 1
\eea
\end{subequations}

The systems \eqref{balance} with parameters given by \eqref{constant} can be resolved recursively for any order and the final solution obtained from the expression \eqref{exp}.
The zeroth order approximation, valid in the limit $K \to \infty$, corresponds to the usual solution found for rate models in the balanced state \citep{bal1,rosenbaum2014},
such solution is physical whenever one of the following inequalities is satisfied
\begin{equation}
\label{ine1}
\frac{I_0^{(e)}}{I_0^{(i)}}>\frac{g_0^{(ei)}}{g_0^{(ii)}}>\frac{g_0^{(ee)}}{g_0^{(ie)}}
\quad , \quad
\frac{I_0^{(e)}}{I_0^{(i)}}<\frac{g_0^{(ei)}}{g_0^{(ii)}}<\frac{g_0^{(ee)}}{g_0^{(ie)}}  \qquad ;
\end{equation}
which ensure the positive sign of $\overline{R}^{(e)}_0$ and $\overline{R}^{(i)}_0$.
The zeroth order solution does not depend on the structural heterogeneity,
since the ratio $\Delta^{(\alpha \alpha)}/K $  vanishes in the limit $K \to \infty$. It should be
stressed that this ratio does not correspond to the coefficient of variation introduced in \citep{landau2016}
to characterize the in-degree distribution. This because we are considering a Lorentzian distribution,
where the average and the standard deviation are not even defined. Moreover, already the first order
corrections depends on $\Delta^{(\alpha \alpha)}_0$.

In order characterize the level of balance in the system one usually estimates the values of the effective input currents 
$i^{(e)}_{eff,j}$ and $i^{(i)}_{eff,j}$ driving the neuron dynamics. These at a population level can be rewritten as
\begin{equation}\label{Ieff}
I_{eff}^{(e)} = \sqrt{K} \left[ I_0^{(e)} + \tau_m(g_{0}^{(ee)} {R}^{(e)}-g_{0}^{(ei)} {R}^{(i)}) \right]
\enskip , \enskip
I_{eff}^{(i)} = \sqrt{K} \left[ I_0^{(i)} + \tau_m(g_{0}^{(ie)} {R}^{(e)}-g_{0}^{(ii)} {R}^{(i)} ) \right]  
\end{equation}
In a balanced state these quantities should not diverge with $K$, instead they should approach
some constant value. In or MF formulation, we can estimate analytically 
the values of the effective currents in the limit $K \to \infty$
for an asynchronous state and they read as
\begin{equation}\label{Ieff_a}
I_{a}^{(e)} = \tau_m \left[ g_{0}^{(ee)} \overline{R}_1^{(e)}-g_{0}^{(ei)} \overline{R}_1^{(i)} \right]
\quad , \quad
I_{a}^{(i)} = \tau_m \left[ g_{0}^{(ie)} \overline{R}_1^{(e)} -g_{0}^{(ii)} \overline{R}_1^{(i)} \right] \quad .
\end{equation}
It should be noticed that these asymptotic values depend on the first order corrections to the balanced solution \eqref{balance}.
Therefore, they depend not only on the synaptic couplings $g_0^{(\alpha \beta)}$ and on the external 
DC currents, but also on the parameters $\Delta_0^{(\alpha \alpha)}$ controlling the structural heterogeneities.

Depending on the parameter values, the currents $I_a^{(\alpha)}$ can be positive or negative, thus
indicating a balanced dynamics where most part of the neurons are supra or below threshold, respectively.
Usually, in order to obtain a stationary state characterized by a low rate and a Poissonian
statistic, as observed in the cortex, one assumes that the excitation and inhibition nearly cancel.
So that the mean membrane potential remains slightly below threshold, and the neurons
can fire occasionally due to the input current fluctuations \citep{bal1,brunel2000}.
However, as pointed out in \citep{lerchner2006} this is not the only possible scenario for a
balanced state. In particular, the authors have developed a self-consistent MF theory for 
balanced Erd\"os-Renyi networks made of heterogeneous Leaky Integrate-and-Fire (LIF) neurons.
In this context they have shown that Poisson-like dynamics are visible only at
intermediate synaptic couplings. While mean driven dynamics are expected for low couplings, 
and at large couplings bursting behaviours appear in the balanced network. Recently,
analogous dynamical behaviours have been reported also for a purely inhibitory heterogeneous 
LIF network \citep{angulo2017}. These findings are consistent with the results 
in \citep{lerchner2006}, where the inhibition is indeed predominant in the balanced
regime.

 \vspace{1.5cm}
\subsubsection{Lyapunov analysis}
\vspace{0.5cm}

To analyse the linear stability of generic solutions of Eqs. \eqref{eq:5a}, we have estimated the corresponding Lyapunov spectrum (LS) $\{ \lambda_k \}$ \citep{lyapunov2016}.
This can be done by considering the time evolution of the tangent vector ${\bf \delta} = \left\{\delta {R}^{(e)}, \delta V^{(e)}, \delta R^{(i)}, \delta V^{(i)} \right\}$,
that is ruled by the linearization of the Eqs.\eqref{eq:5a}, namely
\begin{subequations}\label{tangent}
\begin{eqnarray}
\tau_m \delta \dot{R}^{(e)} &=& \left[2V^{(e)}+g_{0}^{(ee)}\frac{\Delta_{0}^{(ee)}}{\pi}\right] \delta R^{(e)} + 2 R^{(e)} \delta V^{(e)} \\  
\tau_m \delta \dot{V}^{(e)} &=& 2 V^{(e)} \delta V^{(e)} -2 (\pi \tau_{m})^2 R^{(e)} \delta R^{(e)}
+ \sqrt{K} \tau_m \left[ g_{0}^{(ee)} \delta R^{(e)}-g_{0}^{(ei)} \delta R^{(i)} \right] \\
\tau_m \delta \dot{R}^{(i)} &=& \left[2V^{(i)}+g_{0}^{(ii)}\frac{\Delta_{0}^{(ii)}}{\pi}\right] \delta R^{(i)} 
+2 R^{(i)} \delta V^{(i)} \\  
\tau_m \delta \dot{V}^{(i)} &=& 2 V^{(i)} \delta V^{(i)} -2 (\pi \tau_{m})^2 R^{(i)} \delta R^{(i)} + \sqrt{K} \tau_m \left[g_{0}^{(ie)} \delta R^{(e)}-g_{0}^{(ii)} \delta R^{(i)}\right] \quad .
\end{eqnarray}
\end{subequations}

In this case, the LS is composed by four Lyapunov exponents (LEs) $\left\{\lambda_k  \right\}$ with $k=1,\dots,4$, which quantify the average growth rates of infinitesimal perturbations along the orthogonal manifolds. The LEs can be estimated as follows
\begin{equation}
\lambda_k = \lim_{t \to \infty} \frac{1}{t} \log{\frac{|{\bf \delta}_k (t)|}{|{\bf \delta}_k (0)|}}
\quad ,
\end{equation}
where the tangent vectors ${\bf \delta}_k$ are maintained ortho-normal during the time evolution by employing a standard technique introduced in \citep{benettin1980}. The autonomous system will be chaotic for $\lambda_1 > 0$, while a periodic (two frequency quasi-periodic) dynamics will be characterized by $\lambda_1=0$ ($\lambda_1=\lambda_2=0$) and a fixed point by $\lambda_1 <0$. 
  
In order to estimate the LS for the neural mass model we have integrated the direct and tangent space evolution with a  Runge-Kutta 4th order integration scheme with $dt=0.01$ ms,
for a duration of 200 s, after discarding a transient of 10 s.

 \vspace{1.5cm}
\subsubsection{Linear Stability of Stationary Solutions}
\vspace{0.5cm}

The linear stability of the stationary solutions $\{\overline{V}^{(e)}, \overline{V}^{(i)}, \overline{R}^{(e)}, \overline{R}^{(i)}\}$  can be analyzed by solving the eigenvalue problem for the linear equations \eqref{tangent} estimated for stationary values of the mean membrane potentials and of the population firing rates. This approach gives rise to a fourth order characteristic polynomial of the complex eigenvalues  $\Lambda^{(k)} = \Lambda_R^{(k)} + i \Lambda_I^{(k)}$ with $k=1,\dots,4$. The stability of the fixed point is controlled by the maximal $\Lambda_R^{(k)}$, whenever it is positive (negative) the stationary solution is unstable (stable). The nature of the fixed point is determined by $\Lambda_I^{(k)}$, if the imaginary parts of the eigenvalues are all zero we have a node, otherwise a focus. Due to the fact that the coefficients of the characteristic polynomial are real the eigenvalues are 
real or if complex they appear in complex conjugates couples $\Lambda_R^{(j)} \pm i \Lambda_I^{(k)}$. Therefore the relaxation towards the fixed point is characterized by one or two frequencies $\nu_k = \Lambda_I^{(k)}/ (2 \pi)$. These latter quantities, as discussed in details in the following, can give good predictions for the frequencies $\nu_{CO}$ of 
fluctuation driven COs observable for the same parameters in the network dynamics.

In the limit $K >> 1$, we can approximate the linear stability equations \eqref{tangent} as follows:
\begin{subequations}\label{tangent_K}
\begin{eqnarray}
\tau_m \delta \dot{R}^{(e)} &=&  2 \overline{R}^{(e)}_0 \delta V^{(e)} \\  
\tau_m \delta \dot{V}^{(e)} &=& \sqrt{K} \tau_m \left[ g_{0}^{(ee)} \delta R^{(e)}-g_{0}^{(ei)} \delta R^{(i)} \right] \\
\tau_m \delta \dot{R}^{(i)} &=& 2 \overline{R}^{(i)}_0 \delta V^{(i)} \\  
\tau_m \delta \dot{V}^{(i)} &=& \sqrt{K} \tau_m \left[g_{0}^{(ie)} \delta R^{(e)}-g_{0}^{(ii)} \delta R^{(i)}\right] \quad ;
\end{eqnarray}
\end{subequations}
where we have considered the zeroth order approximation for the population rates
$\overline{R}^{(e)}_0$ and $\overline{R}^{(i)}_0$.

In this case the complex eigenvalues $\Lambda^{(k)}$ are given by the following expression:
\begin{equation}
\left[\Lambda^{(k)}\right]^2  = \frac{\sqrt{K}}{\tau_m} \left[ \left( g_0^{(ee)} \overline{R}^{(e)}_0 - g_0^{(ii)} \overline{R}^{(i)}_0 \right)
\pm \sqrt{\left( g_0^{(ee)} \overline{R}^{(e)}_0 + g_0^{(ii)} \overline{R}^{(i)}_0 \right)^2 - 4 g_0^{(ei)} g_0^{(ie)} \overline{R}^{(e)}_0 \overline{R}^{(i)}_0} \right] \quad .
\label{eigen}
\end{equation}
From \eqref{eigen} it is evident that $\Lambda^{(k)} \propto (K)^{1/4}$ and by assuming $I_0^{(i)} \propto I_0^{(e)}$, as we will do in this paper, we also have that $\Lambda^{(k)} \propto (I_0^{(e)})^{1/2}$. Therefore for a focus solution we will have the following scaling relation for the relaxation frequencies for sufficiently large $K$
\begin{equation}
\nu_k^{R} = \frac{\Lambda_I^{(k)}}{2 \pi} \propto \sqrt{I_0^{(e)} K^{1/2}}  \quad ;
\label{scaling}
\end{equation}
this scaling is analogous to that found for purely inhibitory QIF networks in \citep{matteo}.
In \citep{bal1} it has been found that the eigenvalues, characterizing the stability of
the asynchronous state, scale proportionally to $\sqrt{K}$, therefore the convergence
(divergence) from the stationary stable (unstable) solution is somehow slower with $K$
in our model. This is due to the presence in our MF of an extra macroscopic variable, the mean membrane potential,
with respect to the usual rate models.

\section{Results}

 %=====================================================================================
\vspace{0.2cm}
\subsection{Phase diagrams}
\label{bif}
\vspace{0.2cm}

\red{In this sub-section we will investigate the possible dynamical regimes emerging in our model
by employing its neural mass formulation. In particular,} the dynamics of the neural mass model \eqref{eq:5a} takes place in a four dimensional space $\left\{{R}^{(e)}, V^{(e)}, R^{(i)}, V^{(i)} \right\}$ and it depends on 9 parameters,
namely on the four synaptic coupling strengths $\left\{g_{0}^{(ee)}, g_{0}^{(ei)},  g_{0}^{(ii)}, g_{0}^{(ie)} \right\}$, the two external stimulation currents $\left\{I_0^{(e)}, I_0^{(i)} \right\}$, the median in-degree $K$ and the HWHM of the two distributions of the in-degrees $\left\{ \Delta_{0}^{(ee)}, \Delta_{0}^{(ii)} \right\}$.

However, in order to reduce the space of parameters to investigate and at the same time to satisfy the inequalities \eqref{ine1}, required for the existence of a balanced state in the
large $K$ limit, we fix the inhibitory DC current as $I_0^{(i)}=I_0^{(e)}/1.02$ and the synaptic couplings as  $g_{0}^{(ee)}=0.27$, $g_{0}^{(ii)}=0.953939$, $g_{0}^{(ie)}=0.3$, and $g_{0}^{(ei)}=0.96286$ analogously to what done in \citep{wolf}.
Therefore we are left with four control parameters, namely $\Delta_{0}^{(ee)}$,  $\Delta_{0}^{(ii)}$, $I_0^{(e)}$, and $K$, 
that we will vary to investigate the possible dynamical states.

%figure 1
\begin{figure}
\centerline{\includegraphics[scale=0.50]{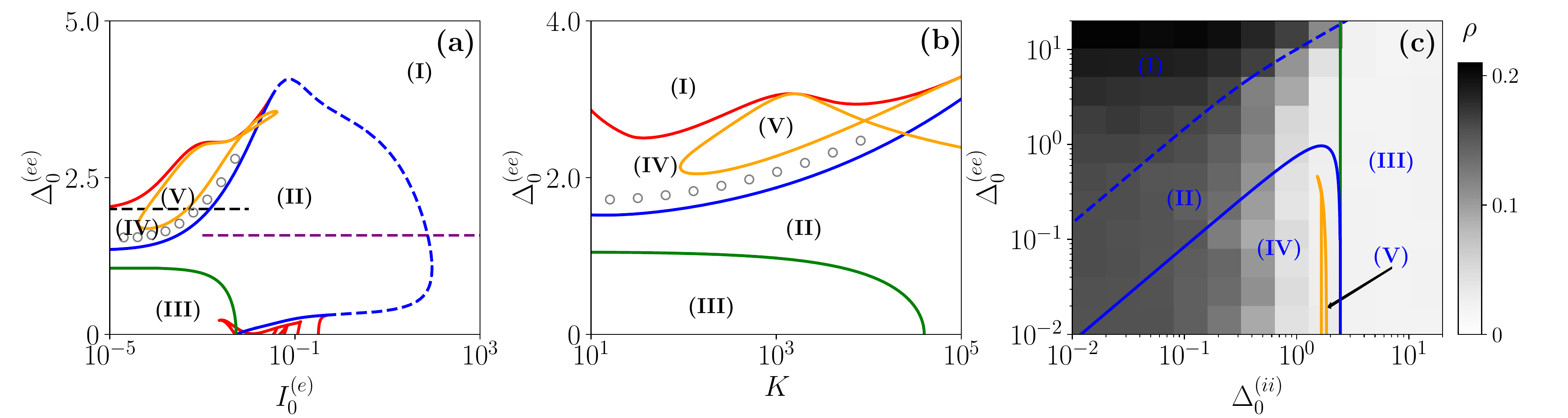}}
\caption{{\bf Bifurcation diagrams of the neural mass model.} The bifurcation diagrams concerns the dynamical state exhibited by the excitatory population in the bidimensional parameter spaces $(I_0^{(e)}, \Delta_0^{(ee)})$ (a),  $(K, \Delta_0^{(ee)})$ (b) and $(\Delta_0^{(ii)}, \Delta_0^{(ee)})$. The regions marked by Roman numbers correspond to the following collective solutions: (I) an unstable focus; (II) a stable focus coexisting with an unstable limit cycle; (III) a stable node; (IV) an unstable focus coexisting with a stable limit cycle; (V) a chaotic dynamics. The green solid line separates the regions with a stable node (III) and a stable focus (II). The blue solid (dashed) curve is a line of super-critical (sub-critical) Hopf bifurcations (HBs), and the red one of Saddle-Node (SN) bifurcations of limit cycles. The yellow curve denotes the period doubling (PD) bifurcation lines. In (c) we report also the coherence indicator $\rho^{(e)}$ \eqref{eq:3}  estimated from the network dynamics with $N^{(e)}=10000$ and $N^{(i)} = 2500$. The dashed lines in (a) indicate the parameter cuts we will consider in Figs. \ref{f4n} and \ref{f5n} (black) and Fig. \ref{f2} (purple),
while the open circles in (a) and (b) denote the set of parameters employed in Fig. \ref{f12}. In the three panels the inhibitory DC current and the synaptic couplings are fixed to $I_0^{(i)}=I_0^{(e)}/1.02$,  $g_{0}^{(ee)}=0.27$, $g_{0}^{(ii)}=0.953939$, $g_{0}^{(ie)}=0.3$, $g_{0}^{(ei)}=0.96286$; other parameters: (a) $K=1000$, $\Delta_{0}^{(ii)}=0.3$ (b) $I_0^{(e)}=0.001$,  $\Delta_{0}^{(ii)}=0.3$ (c) $K= 1000 $ and $I_0^{(e)}= 0.1$.
}
\label{f1}
\end{figure}

Three bidimensional bifurcation diagrams for the neural mass model \eqref{eq:5a} are reported in Fig. \ref{f1} for the couples of parameters $(I_0^{(e)}, \Delta_0^{(ee)})$, $(K, \Delta_0^{(ee)})$ and $(\Delta_0^{(ii)}, \Delta_0^{(ee)})$. From the bifurcation analysis we have identified five different dynamical states for the excitatory population : 
namely, (I) an unstable focus; (II) a stable focus coexisting with an unstable limit cycle; (III) a stable node; (IV)  a stable limit cycle coexisting with an unstable focus;  (V) a chaotic regime. For the analysis reported in the following it is important to remark that the stable foci are usually associated to four complex eigenvalues arranged in complex conjugate couples, therefore the relaxation towards a stable focus is characterized by two frequencies $(\nu_1,\nu_2)$ corresponding to the complex parts of the eigenvalues.
In region (III) the macroscopic fixed point is characterized by two real eigenvalues and a couple of complex conjugated ones. Thus
the relaxation towards the macroscopic node is in this case guided by a single relaxation frequency.
The inhibitory population reveals the same bifurcation structure as the excitatory one, apart an important difference: the inhibitory population never displays stable nodes. Therefore the region (III) for the inhibitory population is also a region of type (II).

As shown in Fig. \ref{f1} (a) and (b), for fixed $\Delta_0^{(ii)}$ and for low values of the structural heterogeneity $ \Delta_0^{(ee)}$ and of the excitatory DC current $I_0^{(e)}$ one observes a stable node (III) that becomes a stable focus (II) by increasing $ \Delta_0^{(ee)}$, these transitions are signaled as green solid lines in Fig. \ref{f1}. By further increasing the degree of heterogeneity $ \Delta_0^{(ee)}$, the stable focus gives rise to collective oscillations (IV) via a super-critical Hopf Bifurcation (HB) (blue solid lines). 
Depending on the values of $K$ and $I_0^{(e)}$ one can have the emergence of chaotic behaviours (V) via a period doubling (PD) cascade (yellow solid lines). For sufficiently large $ \Delta_0^{(ee)}$, the COs disappear via a Saddle-Node (SN) bifurcation of limit cycles (red solid lines) and above the SN line the only remaining solution is an unstable focus (I).

As shown in Fig. \ref{f1} (a), for fixed structural heterogeneities the increase of $I_0^{(e)}$ leads to the disappearance of the stable focus (II) via a sub-critical HB (dashed blue line). The dependence of the observed MF solutions on the in-degree $K$ is reported in Fig. \ref{f1} (b) for a current $I_0^{(e)}=0.001$ and it is not particularly dramatic, apart for the emergence of a chaotic region (V) from a CO regime (IV).

In order to observe the emergence of COs (IV) from the destabilization of a node solution (III) we should vary the structural inhibitory heterogeneity $\Delta_0^{(ii)}$, as shown in Fig. \ref{f1} (c). Indeed, for sufficiently low $\Delta_0^{(ii)}$ and $\Delta_0^{(ee)}$ 
we can observe super-critical bifurcation line  from a node to a stable limit cycle (LC). From this analysis it emerges that the excitatory heterogeneity has an opposite effect with respect to the inhibitory one, indeed by increasing $\Delta_0^{(ee)}$ the value of $\rho^{(e)}$ increases indicating the presence of more synchronized COs. This effect is due to the fact that the increase of $\Delta_0^{(ee)}$ leads to more and more neurons with large $k_j^{(ee)} >> K$, therefore receiving higher and higher levels of recurrent excitation. These neurons are definitely supra-threshold and drive the activity of the network towards coherent behaviours.

In order to understand the limits of our MF formulation, it is of particular interest to compare the network simulations with the MF phase diagram. To this aim, we report in Fig. \ref{f1} (c) also the the coherence indicator $\rho^{(e)}$ \eqref{eq:3}  estimated from the network dynamics. The indicator $\rho^{(e)}$ reveals that no COs are present in region (III), where the MF displays a stable node, however COs emerge in all the other MF regimes for sufficiently low $\Delta_0^{(ii)} < 1$. The presence of COs is expected from the MF analysis only in the regions (IV) and (V), but neither in (II) where the MF forecasts the existence of a stable focus nor in (I) where no stable solutions are envisaged. The origin of the discrepancies among the MF and the network simulations in region (II) is due to the fact that the considered neural mass neglects the dynamical fluctuations in the input currents present in the original networks, that can give rise to noise induced COs \citep{goldobin2021}. However, as shown in \citep{matteo,bi2020} for purely inhibitory populations, the analysis of the neural mass model can still give relevant information on the network dynamics. In particular, the frequencies of the fluctuation induced COs observable in the network simulations can be well estimated from the frequencies $(\nu_1,\nu_2)$ of the relaxation oscillations towards the stable MF focus.
The lack of agreement between MF and network simulations in the region (I) is due to finite size effects,
indeed in this case the system tends to fully synchronize. Therefore, in the network one observes
highly synchronized COs characterized by population firing rates that diverge  for increasing $K$ and $N$
and the MF is unable to reproduce these unrealistic solutions \citep{montbrio2015}.

On the basis of these observations, we can classify the COs observable in the network in three different types accordingly to the corresponding MF solutions: O$_{\rm P}$, when in the MF we observe periodic, quasi-periodic or chaotic collective solutions in regions (IV) and (V);  O$_{\rm F}$, when the MF displays relaxation oscillations towards the stable focus in regions (II) and (III), that in the sparse network become noise sustained oscillations due to fluctuations in the input currents; O$_{\rm S}$, when the MF 
fully synchronizes as in region (I).

\red{
In the following sub-sections we will analyse the macroscopic dynamics of the E-I network of QIF neurons 
in order to test the predictions of the effective neural mass mode for asynchronous as well as coherent dynamics. In this latter case we will focus on the three types of identified COs: namely, O$_{\rm P}$, O$_{\rm F}$ and O$_{\rm S}$. These can manifest as periodic, quasi-periodic and chaotic solutions as we will see by examining two main scenarios indicated as dashed horizontal lines in Fig.  \ref{f1} (a) corresponding to the transition to chaos (black dashed line) and to the emergence of abnormal synchronization from a stable focus (purple dashed line).}

\vspace{1.5cm}
\subsection{Asynchronous Regimes}
\vspace{0.5cm}

\begin{figure}[h!]
\centerline{\includegraphics[scale=0.8]{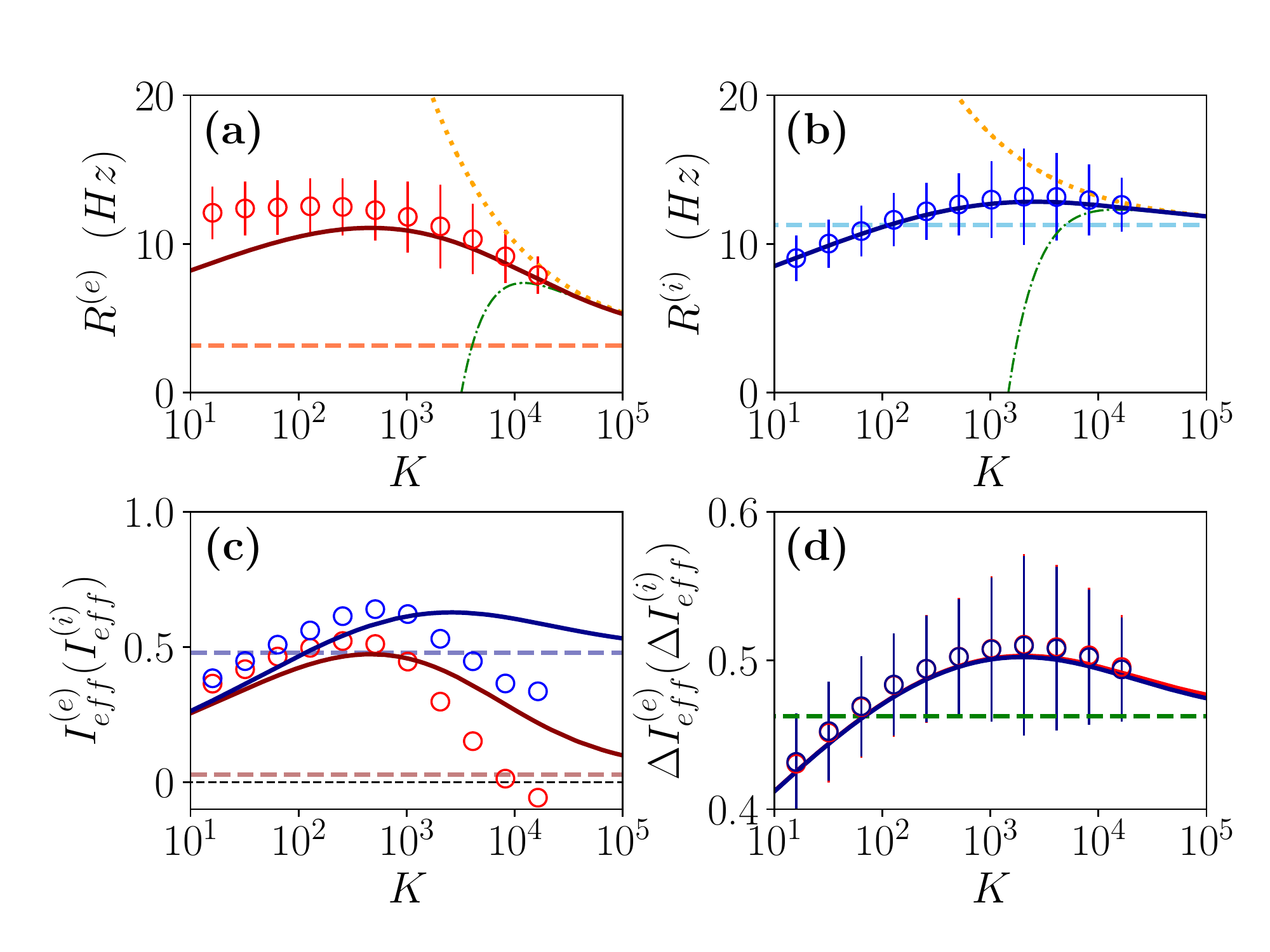}}
\caption{{\bf Asynchronous Dynamics} Instantaneous population rate $R^{(e)}$ ($R^{(i)}$) of excitatory (inhibitory) neurons in function of the median in-degree $K$ are shown in panel (a) (panel (b)).  The effective input currents  $I_{eff}^{(e)}$ ($I_{eff}^{(i)}$) given by Eqs. \eqref{Ieff} are reported in panel (c) and the fluctuations of the input currents 
$\Delta I_{eff}^{(e)}$ ($\Delta I_{eff}^{(i)}$), as obtained  from Eqs. \eqref{fluct}, in panel (d).  
Red (blue) color refer to excitatory) inhibitory population.
The solid continuous lines represent the value obtained by employing the exact MF solutions $\overline{R}^{(x)}$ of \eqref{stat_R}, the dotted (dash-dotted) lines correspond to the first (second) order approximation $\overline{R}^{(x)}_0 + \varepsilon \overline{R}^{(x)}_1$ ($\overline{R}^{(x)}_0 + \varepsilon \overline{R}^{(x)}_1 + \varepsilon^2
\overline{R}^{(x)}_2$) and the dashed horizontal lines to the zeroth order one $\overline{R}^{(x)}_0$ in (a),(b) and (d),
and to $I_a^{(x)}$ in (c)  with $x=e,i$. 
The circles correspond to data obtained from numerical simulations of $N^{(e)}=N^{(i)}=10000$ neurons for $K < 4096$,
$N^{(e)}=N^{(i)}=20000$ for $ K=4096,8192$ and $N^{(e)}=N^{(i)}=30000$ for $K > 8192$,
averaging the population rates over a window of $T=40s$, after discarding a transient of $T=60s$. The error bars in (a) and (b) are obtained as the standard deviations (over the time window $T$) of the population rates, while the average CV of neurons is around 0.15 for all the reported simulations. Synaptic couplings and the ratio between the currents are fixed as stated in sub-section \ref{bif}, other parameters are $\Delta_{0}^{(ii)}=1$ ,  $\Delta_{0}^{(ee)}=2.5$ and $I_0^{(e)}=0.2$. The values of the asymptotic
solutions (dashed lines) are : in (a) and (b) $\overline{R}^{(e)}_0 = 3.18$ Hz and $\overline{R}^{(i)}_0 = 11.28$ Hz, respectively;
in (c) $I_a^{(e)} = 0.0284 $ and $I_{a}^{(i)} \simeq 0.4791$; in (d) $\Delta I_{eff}^{(e)} = 0.4623$ and $\Delta I_{eff}^{(i)} = 0.4593$.
}
\label{fig1}
\end{figure}

We will firstly consider a situation where the network dynamics remains asynchronous for any value of the median in-degree $K$, 
this occurs for sufficiently high structural inhibitory heterogeneities $\Delta_0^{(ii)}$ and external DC currents
as shown in Fig. \ref{f1} (b) and (c) for E-I networks and as reported in \citep{matteo}
for purely inhibitory populations. If the population dynamics is asynchronous, we expect that at a MF level the system will converge towards a stationary state corresponding to a stable equilibrium. Therefore we have compared the results of the network simulations with 
the stationary rates $(\overline{R}^{(e)},\overline{R}^{(i)})$ solutions of \eqref{eq:5a}.  As shown in Fig. \ref{fig1} (a) and (b), the macroscopic activity of the excitatory and inhibitory populations is well reproduced by the  fixed point solutions \eqref{stat_R} in a wide range of values of the in-degrees $ 10 \le K \le 10^4$. This is particularly true for the inhibitory population, while at low $K < 100$  the excitatory firing rate is slightly underestimate by the macroscopic solution $\overline{R}^{(e)}$. 
Due to our choice of parameters, the average inhibitory firing rate is definitely larger than the excitatory one for $K > 100$.
This is consistent with experimental data reported for the barrel cortex of behaving mice \citep{gentet2010}
and other cortical areas \citep{mongillo2018}. Moreover, the rates have a non monotonic behaviour with $K$ with a maximum at $K \simeq 450$ ($K \simeq 2500$) for excitatory (inhibitory) neurons.
As expected, the balanced state solutions $\overline{R}^{(e)}_0 = 3.18$ Hz and $\overline{R}^{(i)}_0 \simeq 11.28$ Hz (dashed horizontal lines) are approached only for sufficiently large $K >> 1$. In Fig. \ref{fig1} (a) and (b) are reported also the first (second) order approximation $\overline{R}^{(e)}_0 + \varepsilon \overline{R}^{(e)}_1$ ($\overline{R}^{(e)}_0 + \varepsilon \overline{R}^{(e)}_1 + \varepsilon^2 \overline{R}^{(e)}_2$) given by Eq. \eqref{balance}. These approximations reproduce quite well the complete solutions already at $K \geq 10^4$.

%figure 1

Let us now consider the effective input currents \eqref{Ieff}, these are reported in Fig. \ref{fig1} (c)
versus the median in-degree. As expected, for increasing $K$
 the MF estimations of the effective currents (solid lines) converge to the asymptotic values
$I_{a}^{(e)} \simeq 0.0284$ and $I_{a}^{(i)} \simeq 0.4791$ (dashed lines) for our choice of parameters.
For the excitatory population the asymptotic value of the effective input current is essentially zero, while
for the inhibitory population it is definitely positive. These results suggest that for the considered choice
of parameters the dynamics of both populations will be balanced, since the quantities $I_{a}^{(e)}$ and $I_{a}^{(i)}$ 
do not diverge with $K$, however at a macroscopic level the excitatory population will be at threshold, 
while the inhibitory one will be supra-threshold.
For comparison, we have estimated $I^{(\alpha)}_{eff}$ also from the direct the network simulations (circles)
for $ 16 \le K \le 16384$. These estimations disagree with the MF results already for $K > 1000$. 
This despite the fact that the population firing rates in the network are very well captured by the MF estimations at large $K$, as shown in Fig. \ref{fig1} (a) and (b). These large differences in the effective input currents are clearly the effect of small
discrepancies at the level of firing rates enhanced by the multiplicative factor $\sqrt{K}$ appearing in Eqs. \eqref{Ieff}.
However, from the network simulations we observe that the effective currents approach values smaller than
the asymptotic ones $I_{a}^{(e)}$ and  $I_{a}^{(i)}$ obtained from the the neural mass model.
In particular, despite the fact that from finite $K$ simulations it is difficult to extrapolate
the asymptotic behaviours, it appears that $I^{(e)}_{eff}$ approaches a small negative value for $K >> 1$,
while  $I^{(i)}_{eff}$ converges to some finite positive value. In the following we will see the effect of these
different behaviours on the microscopic dynamics. The origin of the reported discrepancies 
should be related to the presence of current fluctuations in the network that are neglected in the MF formulation.   
  
The relevance of the current fluctuations for the network dynamics can be appreciated by
estimating their amplitudes within a Poissonian approximation, as follows
\begin{equation}\label{fluct}
\Delta I_{eff}^{(e)} = \sqrt{\tau_m \left[ \left(g_{0}^{(ee)}\right)^2 {R}^{(e)} + \left(g_{0}^{(ei)}\right)^2 {R}^{(i)} \right] } 
\ , \
\Delta I_{eff}^{(i)} = \sqrt{\tau_m \left[ \left(g_{0}^{(ie)}\right)^2 {R}^{(e)} +
\left(g_{0}^{(ii)}\right)^2 {R}^{(i)} \right] }  
\end{equation}
These have been evaluated by assuming that each neuron receive on average $K$ excitatory and inhibitory spike trains 
characterized by a Poissonian statistics with average rates ${R}^{(e)}$ and ${R}^{(i)}$. 
However, we have neglected in the above estimation the variability of the in-degrees of each neuron.
As shown in Fig. \ref{fig1} (d), these fluctuations are essentially identical for excitatory and inhibitory neurons and 
coincide with the MF results. In the limit $K >>1$ they converge to the asymptotic values 
$\Delta I_{eff}^{(e)} \simeq 0.4623$ and $\Delta I_{eff}^{(i)} \simeq 0.4593$ (green dashed lines).
It is evident that already for $K > 1000$ the amplitudes of the fluctuations are of the same order or larger than the effective
input currents. Thus suggesting that the fluctuations have indeed a relevant role in determining the network dynamics
and that one would observes Poissonian or sub-Poissonian dynamics for the neurons, whenever
$I_{a}^{(\alpha)}$ is sub-threshold or supra-threshold \citep{lerchner2006}.

\begin{figure}[h!]
\centerline{\includegraphics[scale=0.5]{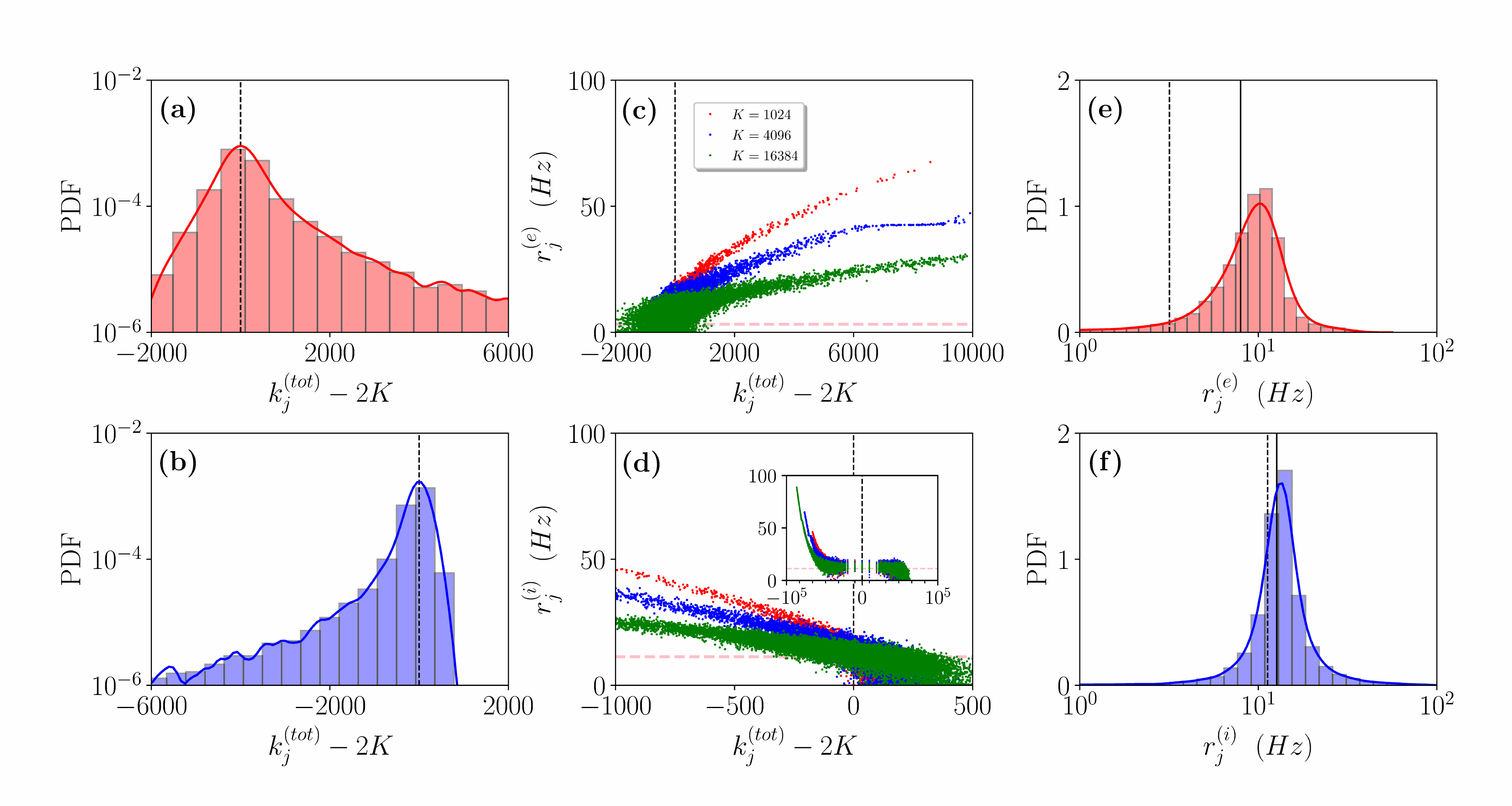}}
\caption{{\bf Asynchronous Dynamics} Probability distribution functions (PDFs) of the total in-degrees $k^{(tot)}_j$ for excitatory (a) and inhibitory (b) active neurons for $K=16384$. (c-d) Firing rates of the excitatory (inhibitory) neurons $r_j^{(e)}$ ($r_j^{(i)}$) versus their total  in-degrees $ k^{(tot)}_j - 2K$ symbols refer to $K=1024$ (red), $K=4096$ (blue) and $K=16384$ (green). The inset in (d) is an enlargement of the panel displaying  the firing rates over the entire scale $ k^{(tot)}_j - 2K$. The magenta dashed lines in (c-d) represent the  balanced state solution $(\overline{R}^{(e)}_0,\overline{R}^{(i)}_0)$. (e-f) PDF of the excitatory (inhibitory) 
firing rates $r^{(e)}_j$ ($r^{(i)}_j$) for $K=16384$, the solid (dashed) line refers to the MF results $\overline{R}^{(x)}$
($\overline{R}^{(x)}_0$) with $x=e,i$. The red (blue) solid line refers to a log-normal fit to the excitatory (inhibitory) PDF with mean $8.8$ Hz ($17.5$ Hz) and standard deviation $3.8$ Hz ($2.3$ Hz). 
%In the inset in panel (e) (panel (f)) the PDF of the effective input currents $i^{(e)}_{eff,j}$ ($i^{(e)}_{eff,j}$) for the excitatory (inhibitory) neurons are displayed for $K=16384$, the solid (dashed) line is the MF result for the effective current at finite $K$, obtained by employing $\overline{R}^{(x)}$, (in the limit $K \to \infty$, corresponding to $I_a^{(x)}$,) with $x=e,i$. 
The  parameters are the same as in Fig. \ref{fig1}, the firing rates have been estimated by simulating the networks for a total time $T_s = 60$ s, after discarding a transient $T_t = 40$ s.}
\label{fig1a}
\end{figure}

In order to understand how the in-degree heterogeneity influences the network dynamics at a microscopic
level we examine the dynamics of active neurons in function of their total in-degree $k^{(tot)}_j$.
This is defined for excitatory (inhibitory) neurons as 
$k^{(tot)}_j = k^{(ee)}_j + k^{(ei)}_j$ ($k^{(tot)}_j = k^{(ii)}_j + k^{(ie)}_j$). Furthermore,
a neuron is considered as active if it has fired at least once during the whole simulation time $T_t + T_s =100$ s,
therefore if it has a firing rate larger than 0.01 Hz. As shown in Figs. \ref{fig1a} (a) and (b), the PDF of active neurons is skewed towards values $k^{(tot)}_j >  2K$ ($k^{(tot)}_j <  2K$) for excitatory (inhibitory) neurons. These results reflect the fact that the excitatory (inhibitory) neurons with low (high) recurrent in-degrees  $k^{(ee)}_j << K$  ($k^{(ii)}_j >> K$) are driven below threshold by the inhibitory activity, that is predominant in the network since $R^{(i)} > R^{(e)}$, $g_0^{(ei)} > g_0^{(ee)}$, and  $g_0^{(ii)} > g_0^{(ie)}$. The number of silent neurons 
for $K > 1024$ is of the order of 6-10 \% for both inhibitory and excitatory populations, in agreement with experimental
results for the barrel cortex of mice \citep{o2010}, where a fraction of 10 \% of neurons was identified as silent with a firing rate
slower than $0.0083$ Hz. It should be remarked that all the population averages we report include the silent neurons.

Let us now examine how the firing rates of active neurons will modify by increasing the value of the median in-degree $K$. The single neuron firing rates as a function of their total in-degrees $k^{(tot)}_j$ are reported in Figs. \ref{fig1a} (c) and (d) for $K=1024,4096$ and 16384. A common characteristics is that the bulk neurons, those with $k^{(tot)}_j \simeq 2K$, tend to approach the firing rate values $(\overline{R}^{(e)}_0,\overline{R}^{(i)}_0)$ (magenta dashed lines) corresponding to the expected solutions for a balanced network in the limit $N >> K \to \infty$ \citep{van1996}. This is confirmed by the analysis of their coefficient of variations $cv_j$, whose values are of order one, as expected for fluctuation driven dynamics. On the other hand, the outlier neurons, i.e. those with $k^{(tot)}_j$ 
far from $2K$, are all characterized by low values of the coefficient of variation $cv_j$ indicating a mean driven dynamics. However, there is striking difference among excitatory and inhibitory neurons. For the excitatory ones we observe that the firing rates of the outliers with $k^{(tot)}_j >> 2 K$  decrease for increasing $K$, while for the inhibitory population the increase of $K$ leads to the emergence of outliers at $k^{(tot)}_j << 2K$ with higher and higher firing rates (see the inset in Fig.\ref{fig1a} (d)). This difference can be explained by the different values measured for $I_{eff}^{(e)}$ and $I_{eff}^{(i)}$ in the network (see Fig. \ref{fig1} (c)).  The increase of $K$ leads for the excitatory (inhibitory) population to the emergence of neurons with very large $k^{(ee)}_j >> K$ (very small $k^{(ii)}_j << K$) whose dynamics should be definitely supra-threshold. 
However, this is compensate in the excitatory case by the rapid drop of $I_{eff}^{(e)}$ towards zero or negative values,
while for the inhibitory population $I_{eff}^{(i)}$ remains positive even at the largest $K$ we have examined.

These outliers seem to have a negligible influence on the population dynamics, as suggested by the fact that
the mean firing rates are well reproduced by the balanced solutions $\overline{R}^{(e)}_0$ and $\overline{R}^{(i)}_0$
and as confirmed also by  examining
the PDFs of the firing rates for $K=16384$. As shown in Figs. \ref{fig1a} (e) and (f), the excitatory (inhibitory) PDF
can be well fitted by a log-normal distribution with mean $8.8$ Hz ($17.5$ Hz) and standard deviation $3.8$ Hz ($2.3$ Hz).
\red{This is considered a clear indication that the network dynamics is fluctuation driven \citep{roxin2011}
as confirmed by recent investigations in the hippocampus and in the cortex \cite{wohrer2013, buzsaki2014, mongillo2018},
as well as in the spinal  motor networks \citep{petersen2016}. However, the relative widths of our distributions are narrower than those reported in \citep{mongillo2018}.
This difference can find an explanation in the theoretical analysis reported in \citep{roxin2011}, where the authors
have shown that quite counter intuitively a wider distribution of the synaptic heterogeneities can lead to a narrower
distribution of the firing rates. Indeed, here we consider Lorentzian distributed in-degrees, while in \citep{mongillo2018}
Erd\"os-Renyi networks have been analyzed.}
As a further aspect, we have estimated the number of inhibitory neurons firing faster than a certain threshold $\nu_{th}$, this number does not depend on the median in-degree for sufficiently large $K > 5000$, however it grows proportionally to $N$. In the considered cases, the fraction of these neurons is $\simeq 1 \%$ for $\nu_{th} = 50$ Hz.

\red{From this analysis we can conclude that at any finite $K$ and for finite observation times we have
at a macroscopic scale an essentially balanced regime sustained by the bulk of active neurons, whose dynamics is fluctuation driven.
Furthermore, we also have a large body of silent neurons as well as a small fraction of mean driven outliers. 
These should be considered as typical features of 
finite heterogeneous neural circuits as shown in various experiments \citep{o2010,landau2016}.
Moreover, in the present case we report a quite different behaviours for outliers whose
macroscopic effective input currents are supra- or sub-threshold. 
}

\vspace{0.5cm}

\subsection{Collective Oscillations}

\vspace{0.5cm}

We will now characterize the different type of COs observable by firstly following a route to coherent chaos for the E-I balanced network and successively we will examine how oscillations exhibiting an abnormal level of synchronization,
somehow similar to those observable during an ictal state in the brain  \citep{lehnertz2009}, can emerge in our system. Furthermore, we will consider the phenomenon of quasi-periodicity and frequency locking occurring for fluctuation driven oscillations. As a last issue, the scaling  of the frequencies and amplitudes of COs with the in-degree and as a function of the external DC current is reported.

 %=====================================================================================
\vspace{0.5cm}
\subsubsection{A period doubling route to coherent chaos}
\vspace{0.5cm}

\begin{figure*}
\centerline{\includegraphics[scale=0.8]{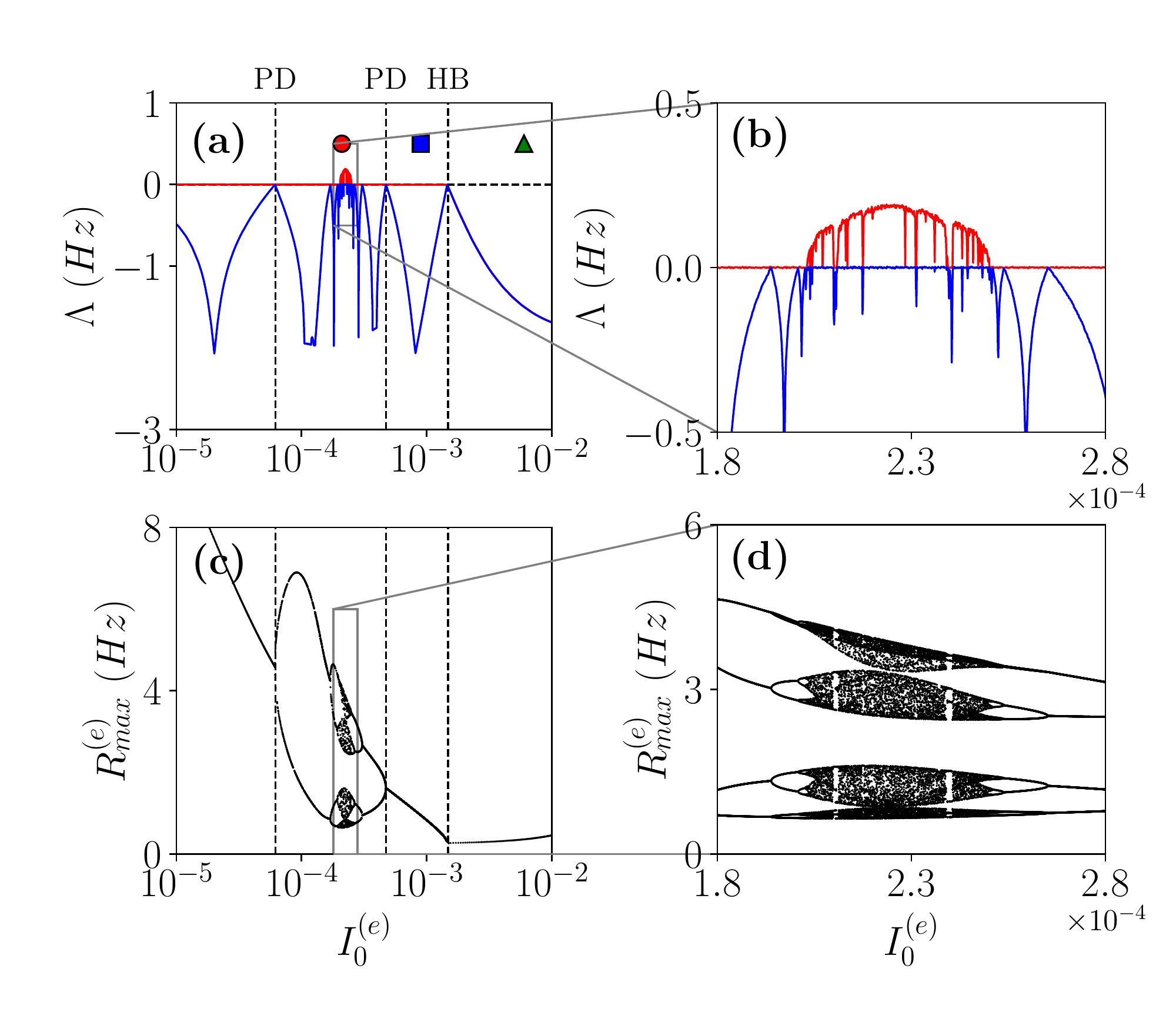}}
\caption{{\bf Coherent Chaos.} (a-b) First (red) $\lambda_1$ and second $\lambda_2$ (blue) Lyapunov exponents for the MF versus the DC current $I_0^{(e)}$ for the parameter cut corresponding to the dashed black line in Fig. \ref{f1} (a). The dashed vertical lines in (a) indicate a super-critical Hopf Bifurcation (HB) from a stable focus to periodic COs and the region of the Period Doubling (PD) cascade. The symbols denote three different types of MF solutions: namely, stable focus (green triangle); periodic oscillations (blue square) and chaotic oscillations (red circle). (c-d) Bifurcation diagrams for the same region obtained by reporting the maximal value of the instantaneous firing rate $R^{(e)}$ measured from MF simulations.
%Here the stable oscillation region is $I_0^e<0.0000617714$ and $0.000568<I_0^e<0.001486$ (grey region). The Multiple-stable oscillations region is $0.0000617714<I_0^e<0.0004729688$ (orange region).  The stable focus region is $0.001486<I_0^e<50.6105$ (lightgrey region). (b,d,f) we show the phase space in $I_0^e=0.00021$ (chaos, $\lambda_1=0.003308, \lambda_2=-0.000021, \lambda_3=-0.080923,  \lambda_4=-0.185512$) (red circle), $I_0^e=0.0009$ (stable oscillation $\lambda_1=0.0, \lambda_2=-0.034325, \lambda_3=-0.055529,  \lambda_4=-0.173195$) (blue square) and $I_0^e=0.006$ (stable focus $\lambda_1=-0.029973, \lambda_2=-0.029948, \lambda_3=-0.101531,  \lambda_4=-0.101538$) (green triangle). (c,e,g) we show raster plot corresponding to (b,d,f). 
The parameters are the same as in Fig. \ref{f1}, other parameters set as $\Delta_0^{(ii)}=0.3$, $\Delta_0^{(ee)}=2.0$, $K = 1000$.
}
\label{f4n}
\end{figure*}

As a first case we will follow the path in the parameter space denoted as a dashed black line in Fig. \ref{f1} (a). In particular, in order to characterize the different dynamical
regimes we have estimated the Lyapunov spectrum $\{\lambda_i\}$ associated to the MF equations. As shown in Fig. \ref{f4n}, this analysis has allowed to identify a period doubling cascade towards a chaotic region, characterized by periodic and chaotic windows. In particular, we observe a focus region (II) for $0.0015 < I_0^{(e)} < 50.6105$,
the focus looses stability via a super-critical Hopf bifurcation at $I_0^e \simeq 0.0015$ giving rise to COs. One observes a period doubling cascade (regime (V)) taking place in the interval $I_0^{(e)} \in [0.00006177 ; 0.00047297]$ followed by a regime of 
COs at lower values of $I_0^{(e)}$. The chaotic dynamics refer to the MF evolution and it can be
therefore definitely identified as collective chaos \citep{nakagawa1993,shibata1998,olmi2011}. A peculiar aspect of this period doubling cascade is that the chaotic dynamics remain always confined in four distinct regions without merging in an unique interval as it happens
e.g. for the logistic map at the Ulam point \citep{ott}. This is due to the fact that the population dynamics displays period four oscillations characterized by four successive bursts, whose amplitudes (measured by $R_{max}^{(e)}$) varies chaotically but each one remains restricted in an interval not overlapping with the other ones.

%figure 4 new

\begin{figure*}
\centerline{\includegraphics[scale=0.65]{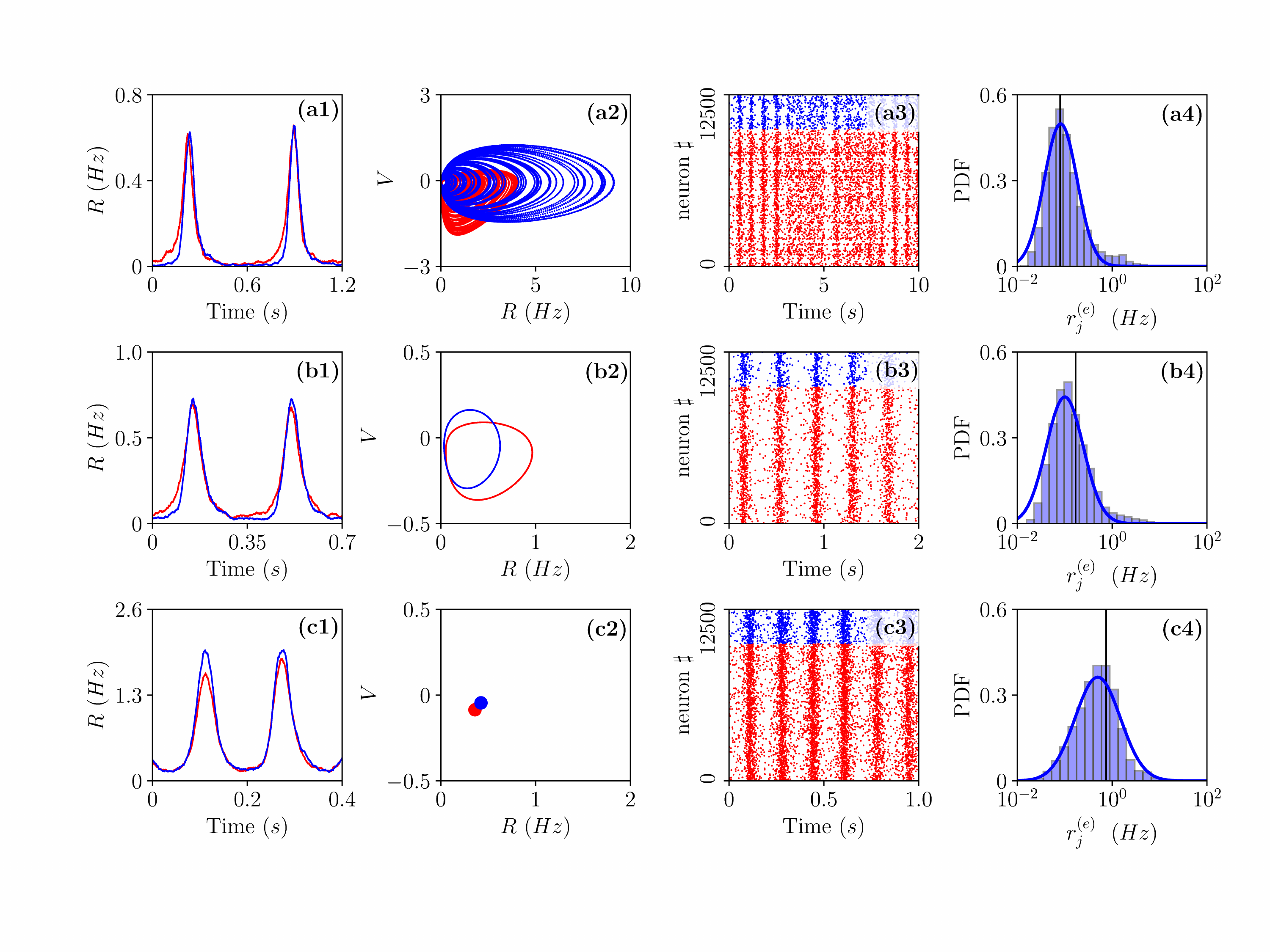}}
\caption{{\bf Different types of collective oscillations.} \red{Row (a) refers to the chaotic state observable for $I_0^{(e)}=0.00021$ in the MF denoted by a red circle in Fig. \ref{f4n} (a); row (b) to the oscillatory state of the MF observable for $I_0^{(e)}=0.0009$ denoted by a blue square in Fig. \ref{f4n} (a); row (c) to the stable focus for the MF observable for
$I_0^{(e)}=0.006$ denoted by a green triangle in Fig. \ref{f4n} (a). The first column displays the population firing rates versus time obtained from the network dynamics, the second the corresponding MF attractors in the planes identified by $(R^{(e)},V^{(e)})$ and $(R^{(i)},V^{(i)})$, the third the raster plots and the fourth
the PDFs of the excitatory firing rates $r_j^{(e)}$. Red (blue) color refers to excitatory (inhibitory) populations,
the solid vertical lines in column 4 to the mean firing rate and the blue solid line to a fit to a log-normal distribution. Parameters as in Fig. \ref{fig1}, apart $\Delta_0^{(ii)}=0.3$, $\Delta_0^{(ee)}=2.0$, $K =1000$. For the estimation of the firing rates we employed $N^{(e)}=40000$ and $N^{(i)}=10000$, while for the raster plots $N^{(e)}=10000$ and $N^{(i)}=2500$. The total integration time has been of 120 sec after discarding a transient of 80 sec.}
}
\label{f5n}
\end{figure*}

Let us now examine the network dynamics for the 3 peculiar MF solutions indicated in Fig. \ref{f4n} (a) corresponding to a stable focus (II) characterized by Lyapunov exponents ($\lambda_1=\lambda_2= -0.0299, \lambda_3=\lambda_4=-0.101$) for $I_0^{(e)}=0.006$ (green triangle), to a stable oscillation (IV) with ($\lambda_1=0.0, \lambda_2=-0.0343, \lambda_3=-0.0555,  \lambda_4=-0.1732$) for $I_0^{(e)}=0.0009$   (blue square), and to collective chaos (v) with ($\lambda_1=0.0033, \lambda_2=0.0, \lambda_3=-0.0809,  \lambda_4=-0.1855$) for  $I_0^{(e)}=0.00021$  (red circle). \red{As shown in Fig. \ref{f5n} for all these three cases the network the dynamics is always characterized by oscillations: namely, O$_{\rm P}$ for the regimes (IV) and (V) and fluctuation induced O$_{\rm F}$ for to the stable MF focus.}

A typical feature of the O$_{\rm P}$ oscillations is that the excitatory neurons start to fire followed by the inhibitory ones, 
furthermore the peak of activity of the excitatory population usually precedes that of the inhibitory neurons of a time interval $\Delta t$. Then the inhibitory burst silences the excitatory population for the time needed to recover 
towards the firing threshold. This recovering time sets the frequency $\nu_{CO}$ of the COs.
In our set-up the excitatory bursts are wider than the inhibitory ones due to the fact that $\Delta_0^{(ee)} > \Delta_0^{(ii)}$. All these features are quite evident from the population firing rates shown in
in Fig. \ref{f5n} (a1) and (b1) and the raster plots in panel (a3) and (b3). These are typical characteristics of a PING-like mechanism reported for the generation of $\gamma$ oscillations in the cortex \citep{tiesinga2009}, despite the fact that the CO's frequencies shown in panels (a) and (d) are of the order of few Hz. Fluctuation driven oscillations O$_{\rm F}$ emerging in the network are radically different, as shown in Fig. \ref{f5n} (c1) in this case the excitatory and inhibitory populations deliver almost simultaneous bursts. \red{Further differences among O$_{\rm P}$ and O$_{\rm F}$ oscillations can be identified at the level
of single neuron activity. These can be appreciated by considering the PDFs of the excitatory firing rates $r_j^{(e)}$ reported in the fourth column of Fig. \ref{f5n}. As shown in Fig. \ref{f5n} (c4) these firing rates are log-normally distributed for O$_{\rm F}$ oscillations, thus confirming their fluctuation driven origin \citep{roxin2011,petersen2016}. On the other hand, for O$_{\rm P}$ oscillations we observe with respect to a log-normal distribution an excess of high firing neurons and a lack of low firing ones
(see Figs. \ref{f5n} (a4) and (b4)). This seems to indicate the presence of a larger number of mean driven excitatory neurons. Indeed this is the case, for $I_0^{(e)}=0.00021$ and $I_0^{(e)}=0.0009$ the percentage of active excitatory neurons driven by average effective currents supra-threshold $i^{(e)}_{eff,j}$ is $\simeq 1.7 - 1.2 \%$, while for $I_0^{(e)}=0.006$ it drops to $\simeq 0.6 \%$. 
The percentage of active inhibitory neurons on average supra-threshold is quite limited in both cases being of the order of
0.25 - 0.13 $\%$. Another interesting feature distinguishing the two kind of oscillations is the fact that for O$_{\rm P}$  the excitatory supra-threshold neurons have a firing rate $r_j^{(e)} > \nu_{CO}$ and that the few neurons with firing rates locked to $\nu_{CO}$ 
are on average exactly balanced, i.e. they have $i^{(e)}_{eff,j} \simeq 0$. The situation is different for the O$_{\rm F}$ oscillations, where we observe a group of sub-threshold excitatory and inhibitory neurons firing locked with the population bursts.
In both cases the most part of neurons are definitely sub-threshold firing at frequencies definitely smaller than $\nu_{CO}$,
as expected for an excitatory-inhibitory balanced network displaying fast network oscillations associated to irregular neural discharges \citep{brunel2003}.}
 
%figure 5 new

\begin{figure*}
\centerline{\includegraphics[scale=0.5]{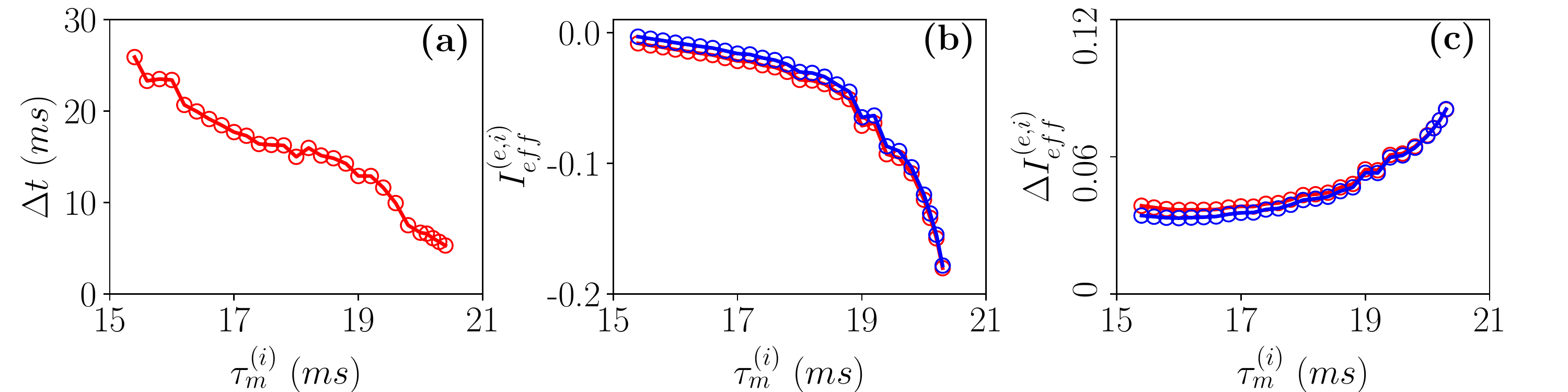}}
\caption{{\bf PING-like O$_P$ collective oscillations.} (a) Firing delays $\Delta t$ between the excitatory population peak and the inhibitory one  versus $\tau_m^{(i)}$. Effective mean input currents \eqref{Ieff} (b) and current fluctuations \eqref{fluct} (c) versus $\tau_m^{(i)}$, the excitatory (inhibitory) population are denoted by red (blue) circles. All the data here reported refer to MF simulations. The parameters are $I_0^{(e)} =0.0009 $,  $\Delta_0^{(ii)}=0.3$, $\Delta_0^{(ee)}=2.0$, $K =1000$, $\tau_m^{(e)}=20$ ms.
}
\label{f6n}
\end{figure*}

In order to understand the different mechanisms at the basis of O$_{\rm P}$ and O$_{\rm F}$ oscillations, let us examine how the delay $\Delta t$ between excitatory and inhibitory bursts, observed for O$_{\rm P}$ oscillations, modifies as a function of the membrane time constant of the inhibitory population $\tau_m^{(i)}$.  An increase of $\tau_m^{(i)}$ of $\simeq 5$ ms has the effect of reducing the delay of almost a factor six from $\Delta t \simeq 28$ ms to $\Delta t \simeq 5$ ms, as shown in Fig. \ref{f6n} (a). The increase of $\tau_m^{(i)}$ leads to an enhanced inhibitory action, since the integration of the inhibitory membrane potentials occurs on longer time scales and this promotes a higher activity of the inhibitory population. Indeed, this is confirmed from the drop of the effective input currents from an almost balanced situation where the average $I^{(e)}_{eff}$ and  $I^{(i)}_{eff}$ are almost zero to a situation where they are definitely negative (see Fig. \ref{f6n} (b)). Thus for increasing $\tau_m^{(i)}$ the percentage of neurons below threshold also increases and as a consequence the dynamics becomes more and more noise driven, as testified by the increase of the current fluctuations $\Delta  I^{(e,i)}_{eff}$ shown in Fig. \ref{f6n} (c). In summary, the delay is due to the fact that, despite the effective inhibitory and excitatory currents are essentially equal, as shown in Fig. \ref{f6n} (b), the wider distribution of the excitatory in-degrees promote the presence of excitatory neurons supra-threshold that are the ones igniting the excitatory burst before the inhibitory one. The delay $\Delta t$ decreases whenever the number of these supra-threshold neurons decreases and it will vanish when the dynamics will become essentially fluctuation driven as in the case of O$_{\rm F}$ oscillations.

%figure 6 new

\vspace{0.5cm}
\subsubsection{From Fluctuation Driven to Abnormally Synchronized Oscillations}
\vspace{0.5cm}

\begin{figure}
\centerline{\includegraphics[scale=0.6]{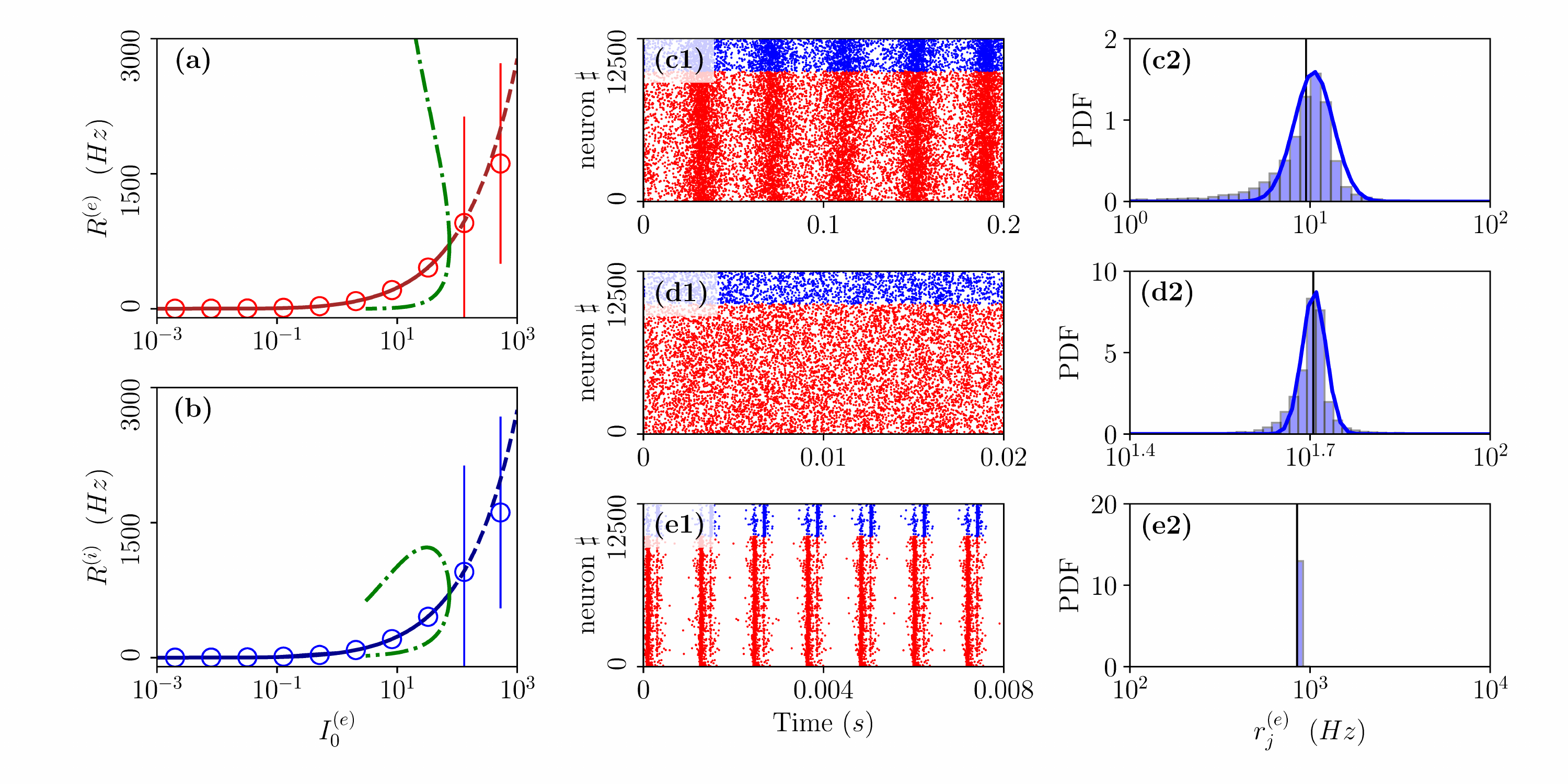}}
\caption{{\bf From fluctuation driven to abnormally synchronized oscillations.}
Firing rates $R^{(e)}$ (a) and $R^{(i)}$ (b) as a function of $I_0^{(e)}$ for E-I networks (circles) and neural mass model (lines) for the parameter cut corresponding to the dashed purple line in Fig. \ref{f1} (a). For the neural mass model: solid (dashed) line shows stable (unstable) focus solution $\overline{R}^{(e)}$ and $\overline{R}^{(i)}$; green dot-dashed lines refer to the extrema of $R^{(e)} (R^{(i)})$ for the unstable limit cycle present in region (II). The unstable limit cycle emerges at the sub-critical Hopf bifurcation for $I_0^{(e)} = 74.1709$ separating region (II) from (I), where the focus becomes unstable. 
\red{Raster plots and PDFs of the excitatory firing rates $r_j^{(e)}$ are reported for specific cases: 
namely, $I_0^{(e)} = 0.128$ (c1,c2) , $I_0^{(e)}=1.024$ (d1,d2) and $I_0^{(e)}=100 $ (e1,e2). The solid vertical lines in (c2,d2,e2) refer to the mean firing rate.} Parameters
as in Fig. \ref{f1}, other parameters are set as $\Delta_0^{(ii)}=0.3$, $\Delta_0^{(ee)}=1.58$, $K =1000$, $N^{(e)}=10000$ and $N^{(i)}=2500$.}
\label{f2}
\end{figure}

As a second range of parameters, we consider the cut in the parameter plane shown in Fig. \ref{f1} (a) as a purple dashed line. For these parameters we report in Fig.  \ref{f2} (a-b) the average in time of the excitatory and inhibitory population rate as a function of the excitatory DC current $I_0^{(e)}$. In particular, we compare network simulations (red and blue circles) with the MF results (red and blue lines). These predict a stable focus (solid lines) up to $I_0^{(e)} =  74.1709$, where a sub-critical Hopf bifurcation destabilizes such solution giving rise to an unstable focus (dashed lines). In panel (a) and (b) we have also reported as green dot-dashed lines the extrema of $R^{(e)}$ and $R^{(i)}$ corresponding to the unstable oscillations emerging at the Hopf bifurcation. We observe a good agreement for the time averaged activity with the MF results for currents smaller than that of the Hopf bifurcation, above which  the MF model predicts a diverging solution. 

\red{In particular, below the Hopf bifurcation, while the MF predicts only the existence of a stable focus the network dynamics reveals quite interesting features. As shown in Fig. \ref{f2} (d1) the system dynamics is indeed asynchronous for intermediate current values, here $I_0^{(e)}=1.024$, however at lower currents we observe fluctuation driven oscillations O$_{\rm F}$ as evident from the raster plot displayed in Fig. \ref{f2} (c1) for $I_0^{(e)} = 0.128$. As shown in Fig. \ref{f2} (c2) and (d2) both these regimes are characterized by log-normal distributions of the firing rates, thus indicating that the dynamics is fluctuation driven.}

As reported in \citep{montbrio2015} when the network dynamics becomes strongly synchronous (as expected for very high common excitatory DC external current) the MF formulation fails since the population rates predicted within the MF formulation diverge. However, as shown in Fig. \ref{f2} (e1,e2) due to finite size effects we observe in the network a strongly synchronous COs of type O$_{\rm S}$ corresponding to the MF region (I) where the MF model predicts no stable solution. These abnormally synchronized oscillations are also characterized by a quite fast frequency of oscillation $\nu_{\rm CO} \simeq 800 - 1000$ Hz. Furthermore, similarly to the O$_{\rm P}$ oscillations they emerge due to a PING-like mechanism. This is evident from the raster plot in  Fig. \ref{f2} (e1), where excitatory neurons fire almost synchronously followed, after an extremely short delay, by the inhibitory ones whose activity silence all the network until the next excitatory burst. Quite astonishingly the mean population rates measured in the network are reasonably well captured by the MF solutions associated to the unstable focus even beyond the Hopf bifurcation, despite the network is now displaying COs (see Fig. \ref{f2} (a-b)).

%figure 2

\begin{figure}
\centerline{\includegraphics[scale=0.8]{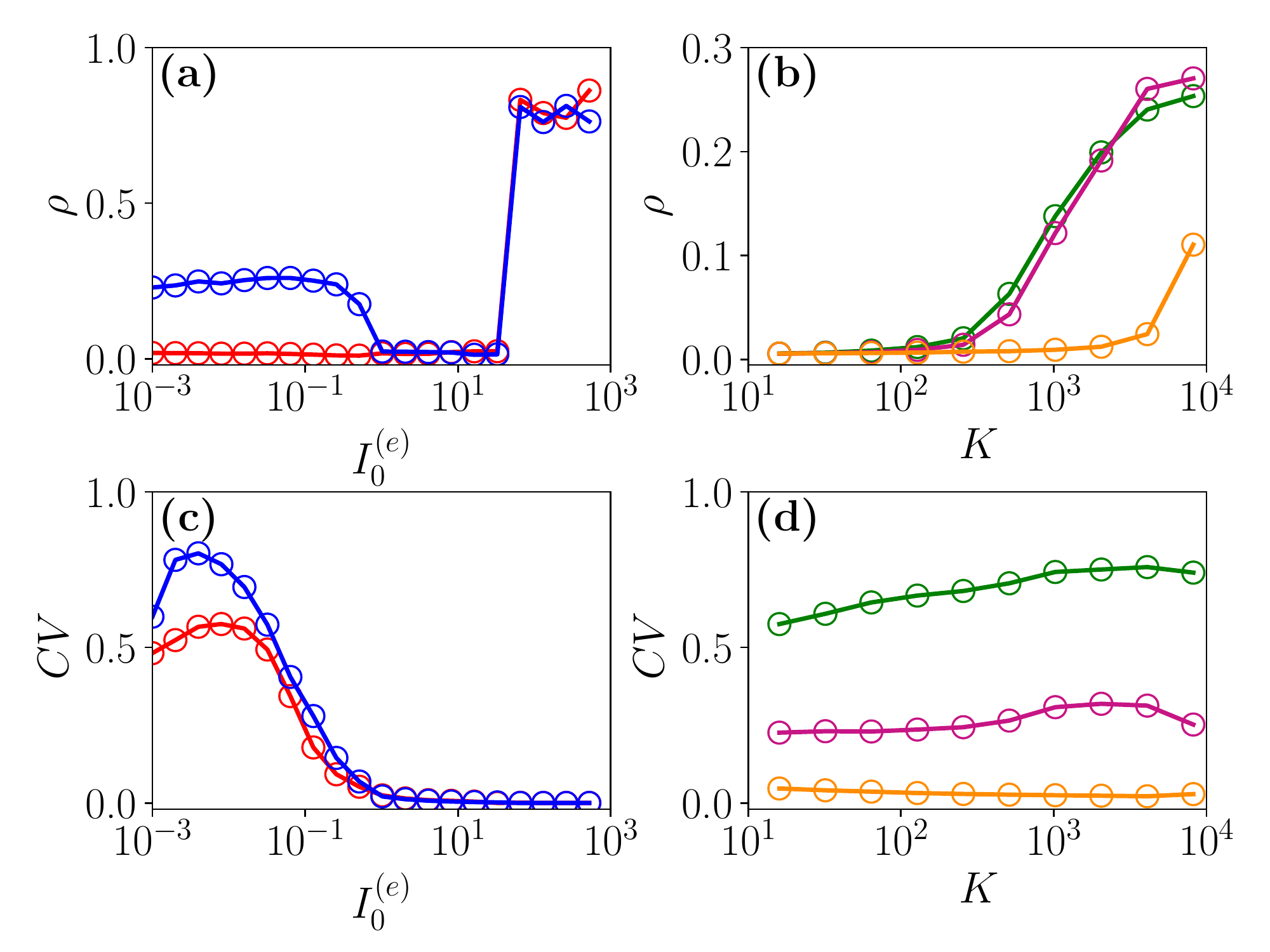}}
\caption{{\bf From fluctuation driven to abnormally synchronized oscillations.}  
Coherence indicator $\rho$ \eqref{eq:3} for the whole network of excitatory and inhibitory neurons versus the excitatory DC current $I_0^{(e)}$ (a) and the median in-degree $K$ (c). Coefficient of variation $CV$ for the whole network versus $I_0^{(e)}$ (b) and $K$ (d). In panel (a) and (c) the symbols refer to different values of the median in-degree:namely,  $K=100$ (red circles) and $K=4000$ (blue circles). In (b) and (d) the symbols refer to different excitatory DC currents: namely,  $I_0^{(e)}=0.01$ (green circles), 
$I_0^{(e)}=0.1$ (purple circles) and $I_0^{(e)}=1.0$ (orange circles).
Parameters as in Fig. \ref{f1}, other parameters $\Delta_0^{(ii)}=0.3$, $\Delta_0^{(ee)}=1.58$, $N^{(e)}=40000$ and $N^{(i)}=10000$.}
\label{f3}
\end{figure}

The emergence of COs in the network can be characterized in terms of the coherence indicator $\rho$ \eqref{eq:3} for the whole population of neurons. This indicator is reported in Fig. \ref{f3} (a) as a function of $I_0^{(e)}$ for the same parameters previously discussed in Fig. \ref{f2} and for two different values of the median in-degree : $K=100$ (red circles) and $K=4000$ (blue circles). For both values of $K$, we observe an almost discontinuous transition in the value of the coherence indicator at the sub-critical Hopf bifurcation from $\rho \simeq 1/\sqrt{N}$, expected for an asynchronous dynamics, to values $\rho \simeq 1$ corresponding to fully synchronization. This discontinuous transition leads to the emergence of abnormally synchronized oscillations O$_{\rm S}$ in the network. Moreover, at sufficiently high in-degrees we observe the emergence of a new coherent state for low DC currents $I_0^{(e)} < 1.024$ characterized by a finite value of the coherence indicator, namely $\rho \simeq 0.3$. The origin of these oscillations can be better understood by examining the 
coefficient of variation $CV$ averaged over the whole population, this is reported in Fig. \ref{f3} (c) for the same interval of excitatory DC current and the same in-degrees as in  Fig. \ref{f3} (a). It is evident that the $CV$ assumes finite values only for small input currents, namely $I_0^{(e)} < 1.024$, indicating the presence of not negligible fluctuations in the network dynamics. Furthermore, by increasing $K$ these fluctuations, as measured by the $CV$, increases as expected for a balanced network. This analysis suggests that these oscillations cannot exist in absence of fluctuations in the network and therefore they are of the O$_{\rm F}$ type. Furthermore, the network should be sufficiently connected in order to sustain these COs, as one can understand from Fig. \ref{f3} (b) and (d), where $\rho$ and $CV$ are reported as a function of $K$ for three different values of $I_0^{(e)}$. Indeed, for these parameter values no O$_{\rm F}$ oscillation is observable for $K < 400$, even in presence of finite values of the $CV$.

As previously discussed in \citep{matteo}, the balance between excitation and inhibition generates endogenous fluctuations that modifies the collective dynamics with respect to that predicted by the MF model, where the heterogeneity of the input currents, due to distributed in-degrees, is taken in account only as a quenched form of disorder and not as a dynamical source of noise. However, also from this simplified MF formulation one can obtain relevant information on the O$_{\rm F}$ oscillations, indeed as we will see in the next sub-section the relaxation frequencies towards the stable MF focus represent a good estimation of the oscillation frequencies measured in the network. This suggests that the fluctuations present at the network level can sustain COs by continuously exciting the focus observed in the effective MF model with quenched disorder.

%figure 3

\vspace{0.5cm}
\subsubsection{Fluctuation driven oscillations: from quasi-periodicity to frequency locking}
\vspace{0.5cm}

%figure 5
\begin{figure}
\centerline{\includegraphics[scale=0.65]{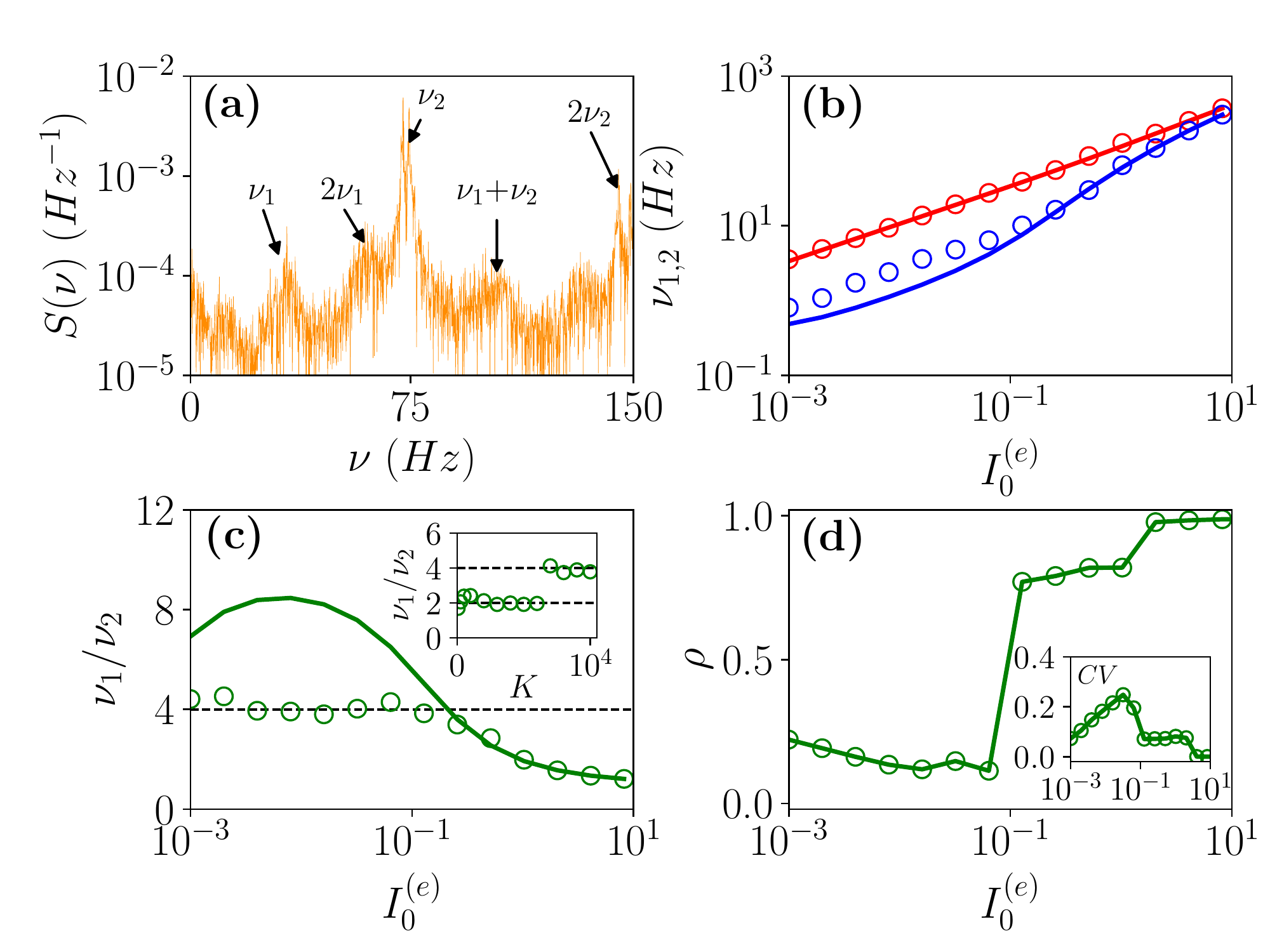}}
\caption{{\bf From quasi-periodicity to frequency locking.}
(a) Power spectra $S(\nu)$ of the mean membrane potential obtained from network simulations. (b) The two fundamental frequencies $\nu_1(\nu_2$) versus $I_0^{(e)}$.
(c) Frequency ratio $\nu_1/\nu_2$ versus $I_0^{(e)}$, in the inset $\nu_1/\nu_2$ is shown versus $K$.
(d) Coherence parameter $\rho$ versus $I_0^{(e)}$, in the inset the corresponding $CV$ is reported.
 In (b-c) the symbols (solid lines) refer to $\nu_1$ and $\nu_2$ as obtained from the peaks of the power spectra $S(\nu)$ for $V(t)$ obtained from the network dynamics (to the the two relaxation frequencies $\nu_1^R$ and $\nu_2^R$ associated to the stable focus solution for the MF). Parameters
as in Fig. \ref{f1}, other parameters are set as $\Delta_0^{(ii)}=0.3$, $\Delta_0^{(ee)}=1.58, N^{(e)}=80000, N^{(i)}=20000$, $K=8192$ and $I_0^{(e)} = 0.128$ in the inset of panel (c).
}
\label{f5}
\end{figure}

As announced, this sub-section will be devoted to the characterization of the fluctuation driven oscillations O$_F$ emerging in region (II) reported in Fig. \ref{f1}. As the MF is now characterized by a stable focus with two couples of complex conjugate eigenvalues there are two frequencies that can be excited by neurons' irregular firing. Accordingly, as reported in \citep{matteo}, we expect the collective dynamics to be characterized by a quasi-periodic dynamics with two (incommensurable) frequencies. These frequencies can be estimated by computing the power spectrum $S(\nu)$ of global quantities, e.g. mean membrane potential  $V(t)$. In the case of a periodic dynamics $S(\nu)$ is characterized by one main peak in correspondence of the CO frequency and minor peaks at its harmonics, while in the quasi-periodic case the power spectrum shows peaks located at the two fundamental frequencies and 
at all their linear combinations. Indeed, as shown in Fig. \ref{f5} (a) the power spectrum exhibit several peaks over a continuous profile and the peak frequencies can be obtained as a linear combination of two fundamental frequencies $(\nu_1,\nu_2)$. As already mentioned, the noisy background  is due to  the fluctuations present in the balanced network.
It is evident from Fig. \ref{f5} (b) that these two fundamental frequencies are well reproduced by the two relaxation frequencies
$\nu_1^R$ and $\nu_2^R$ towards the MF focus, in particular for $I_0^{(e)} \geq 0.256$. At smaller currents, while the first frequency is well reproduced by $\nu_1^R$, the second one is under-estimated by $\nu_2^R$. This is due to the phenomenon of frequency locking among the two collective rhythms present in the system: when the two frequencies become commensurable we observe a common periodic CO.
The locking order can be estimated by plotting the ratio between the two frequencies, indeed for low currents and $K=8192$ the ratio is almost constant and equal to four denoting a 1:4 frequency locking (see  Fig. \ref{f5} (c)). Furthermore, by fixing $I_0^{(e)} = 0.128$ and by varying $K$ the ratio $\nu_1/\nu_2$ can display different
locked states, passing from a locking of type $1:2$ at low $K$ to $1:4$ at larger values, as shown in the inset of  Fig. \ref{f5} (c). 

As evident from Fig. \ref{f5} (b) and (c), the locking phenomenon arises only in the network simulations and it is not captured by the MF model. Furthermore, frequency locking occurs at low currents $I_0^{(e)} < 0.1$  where the dynamics of the neurons is driven by the intrinsic current fluctuations present in the network, but not in the MF. Indeed for low DC currents the level of synchronization within the populations measured by $\rho$ decreases with $I_0^{(e)}$, while the $CV$ increases (as shown in  Fig. \ref{f3} (d)).
These features suggest that this phenomenon is somehow similar to what reported in \citep{meng2018} for two coupled inhibitory neural populations subject to external uncorrelated noise. The authors in \citep{meng2018} observed an increase of the locking region among collective rhythms by increasing the amplitude of the additive noise terms, this joined to a counter-intuitive decrease of the level of synchronization among the neurons within each population. However, in \citep{meng2018} the neurons are subject to independent external noise sources, while in our case the sources of fluctuations are intrinsic to the system and induced by 
the structural heterogeneity. Due to the network sparseness the current fluctuations experienced by each neuron 
can be assumed to be indeed uncorrelated \citep{brunel1999}. Therefore we are facing a new phenomenon that we can identify as a
frequency locking of collective rhythms promoted by self-induced uncorrelated fluctuations. Indeed, the locking disappears for increasing
external DC currents $I_0^{(e)} > 0.1$, when  the coherence parameter $\rho$ displays an abrupt jump towards higher values
and the $CV \simeq 0$, thus indicating that in this regime the neuron dynamics becomes essentially mean driven.

 %=====================================================================================
\vspace{0.5cm}
\subsubsection{Features of COs for large in-degrees and DC currents}
\vspace{0.5cm}

The dynamics of balanced networks is usually characterized in the limit $N >> K >> 1$ by the emergence
of a self-sustained asynchronous regime. However, limit cycle solutions have been already reported for balanced networks in the seminal paper by Van Vreeswijk and Sompolinsky \citep{bal1}. These solutions can be either unbalanced or balanced, however in this latter case they were characterized by vanishing small oscillations' amplitudes. As a matter of fact, the authors in \citep{bal1} have shown that balanced COs are not observable in their model in the limit $N >> K \to \infty$, but only for finite $K$. Therefore, it is important to address also in our case if COs can still be observable in the limit $N >> K >> 1$. \red{Thus, in the following we will investigate the dependence of COs features on the median in-degree $K$ and on the external DC currents.}

 %figure 11
\begin{figure}
\centerline{\includegraphics[scale=0.65]{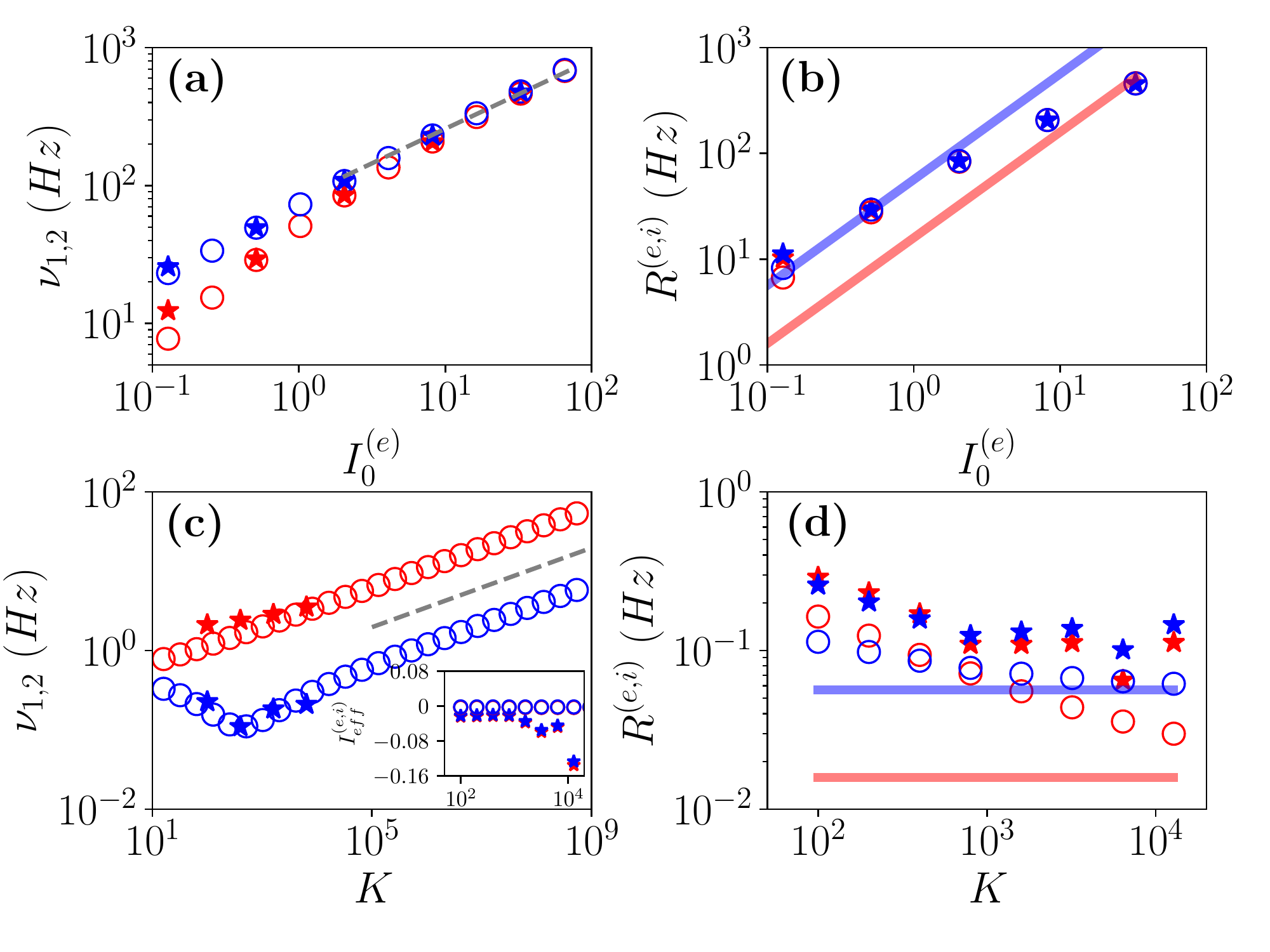}}
\caption{ {\bf Frequencies and amplitudes of O$_F$ oscillations.}
\red{The two fundamental frequencies $\nu_1$ and $\nu_2$ versus $I_0^{(e)}$ (a) and $K$ (c) and the average firing rates
versus versus $I_0^{(e)}$ (b) and $K$ (d)  for the excitatory (red) and inhibitory (blue) populations.
In the inset in (c) the effective mean input currents $I_{eff}^{(e)}$ ($I_{eff}^{(i)}$) of the excitatory
(inhibitory) population are shown versus $K$.  
The dashed line in panel (a) (panel (c)) corresponds to a power law-scaling $\propto {I_0^{(e)}}^{1/2}$ ($\propto K^{1/4}$) for the frequencies of the COs. The solid red (blue) line in panel (b) and (d)
denotes the asymptotic MF result ${\overline R}^{(e)}$ (${\overline R}^{(i)}$).
Network (MF) simulations are denoted as stars (circles). The MF data refer to the stable focus, in
particular in panels (a) and (c) these are the two relaxation frequencies $\nu_1^R$ and $\nu_2^R$.
} Parameters as in Fig. \ref{f1}, other parameters: (a-b) $K=1000$, $\Delta_0^{(ee)}=1.58$, $\Delta_0^{(ii)}=0.3$; (c-d) $I_0^{(e)} = 0.001 $, $\Delta_0^{(ee)}=1.3$, $\Delta_0^{(ii)}=0.3$; for the network simulations we employed $N^{(e)}=80000$ and $N^{(i)}=20000$.}
\label{f11}
\end{figure}

Let us first consider fluctuation driven O$_F$ oscillations, in this case we have an analytical prediction \eqref{scaling} for the scaling of the fundamental frequencies $\nu_k^R$ associated to
the relaxation towards the macroscopic focus, which should grow proportionally to $\sqrt{I^{(e)}}$. As shown in Figs. \ref{f11} (a) and (c), indeed this scaling is clearly 
observable for sufficiently large $K$ and $I_0^{(e)}$. It is also evident the extreme good agreement between results obtained from the network simulations and the theoretical predictions \eqref{scaling}, at least for the values of $K$ reachable with our simulations. Furthermore, the COs' frequencies cover an extremely large range of values from few Hz to KHz and this range of frequencies can be spanned by varying either $K$ or the external DC current $I_0^{(e)}$ as shown in  Fig. \ref{f11} (a) and (c).

\red{To better characterize these regimes we have also evaluated the average firing rates $R^{(e)}$ and $R^{(i)}$.
These quantities are displayed for O$_F$ oscillations in Figs.  \ref{f11} (b) and (d) as a function of $I_0^{(e)}$ and $K$, respectively. From the network simulations (stars) we observe that $R^{(e)}$ and $R^{(i)}$ grow with $I_0^{(e)}$ and
they are astonishingly quite well reproduced by the MF data (circles) for sufficiently large DC currents,
despite the MF results refer to a stable focus and not to COs. Instead, at smaller currents (namely, $I_0^{(e)} = 0.001$)
the network data overestimates the MF results and the excitatory and inhibitory firing rates 
for $K >> 1$ seem to converge to a common constant value definitely larger than those corresponding to the asynchronous regimes.
For sufficiently large $K$, due to the prevalence of inhibition over excitation in the present model we expect that the system will be sub-threshold, since the average excitatory and inhibitory firing rates are essentially coincident. Indeed this is confirmed
by the analysis of the mean effective input currents $I_{eff}^{(e)}$ and $I_{eff}^{(i)}$ shown in the inset of Figs.  \ref{f11} (c).
While for the MF focus the dynamics appear as almost exactly balanced for all
the considered median in-degree $K$ since $I_{eff}^{(e)} \simeq I_{eff}^{(i)} \simeq 0$, for the network dynamics 
$I_{eff}^{(e)}$ and $I_{eff}^{(i)}$ are definitely negative for $K > 1000$.  This does not prevent the emergence of COs driven by fluctuations at large $K$, as indeed observed.
}

\red{These results seem to indicate that for $N >> K \to \infty$ the network will not converge in this case towards a balanced regime
characterized by constant effective input currents. On the contrary from our analysis it emerges
that the system will become more and more sub-threshold for increasing $K > 1000$. However, the system always exhibits a fluctuation driven dynamics, since we measured $CV \simeq 0.6 - 0.8$ at least in the range $K \simeq 100 - 10^4$ accessible to network simulations. }

 %figure 12
\begin{figure}
\centerline{\includegraphics[scale=0.65]{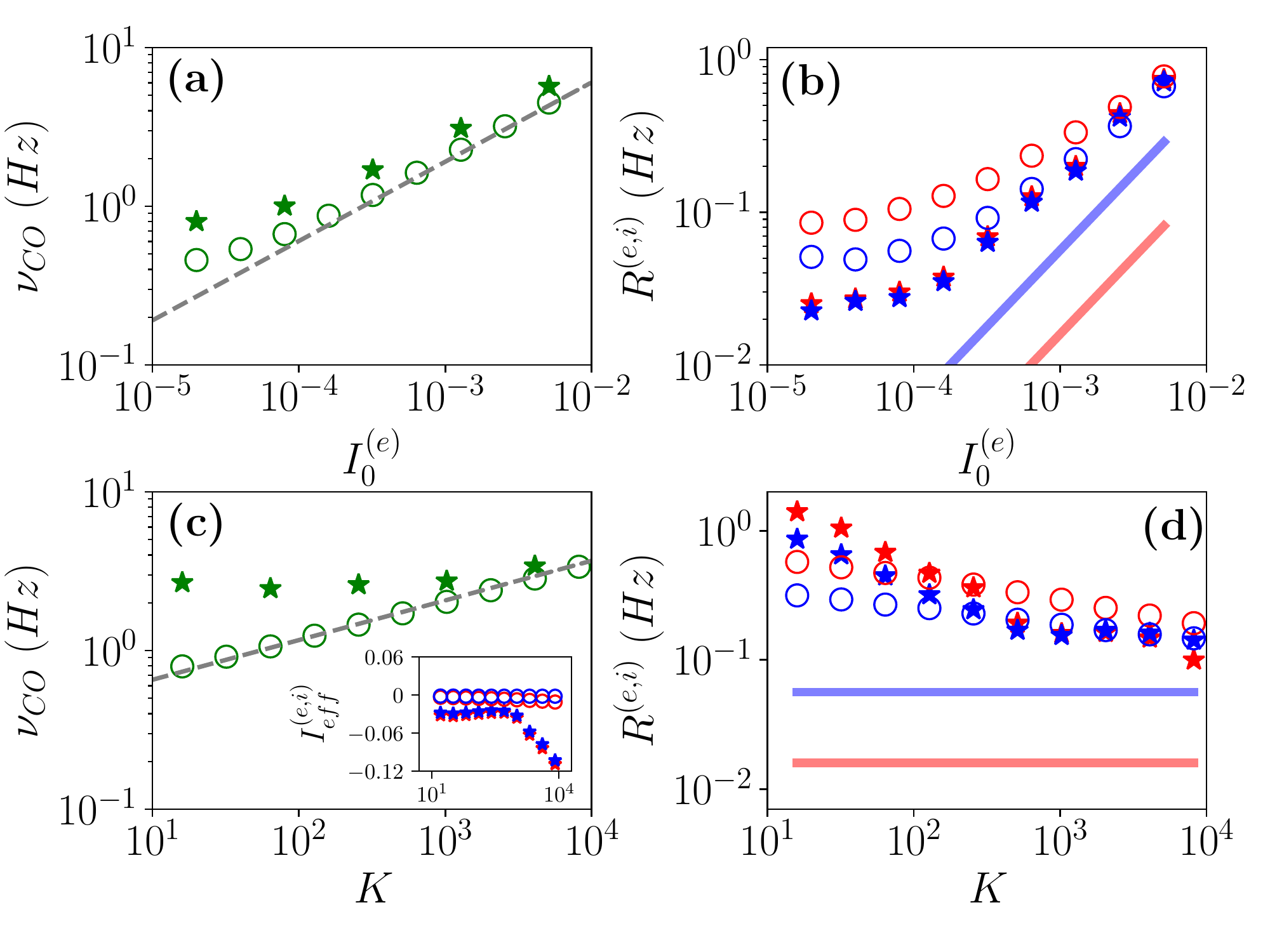}}
\caption{ {\bf Frequencies and amplitudes of O$_P$ oscillations.}
\red{COs' frequency $\nu_{CO}$ versus $I_0^{(e)}$ (a) and $K$ (c)  and mean firing rates versus $I_0^{(e)}$ (b) and $K$ (d)
for the excitatory (red) and inhibitory (blue) populations. The dashed line in panel (a) (panel (c)) corresponds to a power law-scaling $\propto {I_0^{(e)}}^{1/2}$ ($\propto K^{1/4}$) for the frequencies.  In the inset in (c) the effective mean input currents $I_{eff}^{(e)}$ ($I_{eff}^{(i)}$) of the excitatory (inhibitory) population are shown versus $K$. The solid red (blue) line in panel (b) and (d) denotes the asymptotic MF result ${\overline R}^{(e)}$ (${\overline R}^{(i)}$).
} The data obtained from network (MF) simulations are denoted as stars (circles).
The data reported in (a-b) and (c-d) refer to the open circles in Fig. \ref{f1} (a) and (b), respectively. For network simulations we employed $N^{(e)}=80000$ and $N^{(i)}=20000$.}
\label{f12}
\end{figure}

Let us now examine the O$_P$ oscillations. As shown in Figs. \ref{f12} (a) and (c), the frequencies $\nu_{CO}$ as estimated from
the MF model (open circles) reveal an almost perfect increase proportional to $\sqrt{I^{(e)}}$ analogous to the one reported for O$_F$ oscillations. The data obtained from network simulations (stars) converge towards the MF results for sufficiently large $K$ and $I_0^{(e)}$.

\red{The mean firing rates $R^{(e)}$ and $R^{(i)}$ grow with $I_0^{(e)}$ for fixed $K$ and appear to converge
towards a constant value for sufficiently large $K$ for fixed $I_0^{(e)}$, see Figs. \ref{f12} (b) and (d).
Moreover, the network simulations (stars) approach the MF results (open circles) at large DC currents
and median in-degrees.  However, while in the MF the asymptotic values of $R^{(e)}$ and $R^{(i)}$ remain
distinct even at large $K$, these seem to become identical in the network simulations.
This reflects in the fact that while the MF is perfectly balanced in the whole range of examined in-degrees, since $I_{eff}^{(e)} \simeq I_{eff}^{(i)} \simeq 0$ , the network simulations reveal almost balanced effective input currents up to $K \simeq 1000$ and above such median in-degree a prevalence of the inhibitory drive (see inset of Figs. \ref{f12}  (c)).}

 \red{For both kinds of COs we observe that while $\nu_{CO}$ diverges with $K$, the mean firing rates approach a constant
 value, thus suggesting that the percentage of neurons participating to each population burst should vanish in the
 limit $K \to \infty$. This result indicates that COs will finally disappear, however more refined analysis are needed
 to derive the asymptotic behaviour of the system in the large $K$ limit, see \citep{noi} for a detailed
 discussion of this aspect for purely inhibitory networks.
 }

\vspace{0.5cm}
 \section{Discussion}
 
\vspace{0.5cm}

\red{We have extensively characterized the macroscopic regimes emerging in a sparse balanced E-I network made of spiking QIF neurons with
Lorentzian distributed in-degrees. The considered neuronal model joined to the peculiar choice of the distribution
has allowed us to derive an exact low} dimensional neural mass model describing the MF dynamics of the network in terms of the mean membrane potentials and of the population rates of the two populations \citep{montbrio2015, matteo}.
The low-dimensionality of the MF equations enabled us to study analytically the 
stationary solutions and their stability as well as to obtain the bifurcation diagrams associated to the model
and to identify the possible macroscopic states.

\vspace{0.5cm}
\subsection{Asynchronous Regime}
\vspace{0.5cm}

The stationary solutions of the MF correspond to the asynchronous regime, which is the regime usually analyzed
in the context of balanced dynamics \citep{bal1,bal2,bal3}. In the present case we have analytically
obtained the stationary solutions for the mean membrane potentials and average firing rates for Lorentzian distributed in-degrees
for any finite value of the median $K$ and for a HWHM scaling as $\Delta_k^{(\alpha \alpha)} = \Delta_0^{(\alpha \alpha)} {(K)^\eta}$
with $\eta=1/2$. The MF estimations for the population firing rates are pretty well reproduced by the network simulations 
in the examined range of in-degrees $K$. Furthermore, from the analytic expression of the stationary firing rates \eqref{stat_R} 
it is evident that for $K >> 1$ the asymptotic rates would not depend on the structural heterogeneity and correspond to those 
usually found for balanced homogeneous or Erd\"os-Renyi networks \citep{bal1,wolf}. This is due to the fact that the ratio 
$\left(\Delta_k^{(\alpha \alpha)}\right)^2/K$ remains constant for $K \to \infty$. 
The final scenario will depend on the scaling exponent $\eta$, in particular by assuming $\eta = 3/4$
the asymptotic firing rates $\overline{R}_0^{(\alpha)}$ will explicitly depend on the parameters 
$\Delta_0^{(\alpha \alpha)}$ controlling the structural heterogeneity. 
Whenever $\eta > 3/4$ the balanced state breaks down and we face a situation
similar to those investigated in \citep{landau2016,pyle2016} \footnote{\red{In such cases, balance has been recovered
either by rewiring  the post-synaptic connections \citep{pyle2016} or by introducing some sort of homeostatic 
plasticity or of spike-frequency adaptation \citep{landau2016}.}}

However, despite the system approaches a balanced state, as testified by the fact that the effective
input currents converge to finite values $I_a^{(\alpha)}$ \red{and
the current fluctuations stay finite for  $K \to \infty$}, the balanced regime is not necessarily
a sub-threshold one. Indeed, we have observed that we can have either sub-threshold or supra-threshold situations 
depending on the model parameters in agreement with the results previously reported in \citep{lerchner2006}. Moreover,
the excitatory and inhibitory populations can achieve balanced regimes characterized by different asymptotic dynamics,
\red{where $I_a^{(i)}$  and $I_a^{(e)}$ have opposite signs}.
 
While at a macroscopic level the population activity for $N >> K >> 1$ approaches essentially that of
a homogeneous balanced system, as shown in Fig. \ref{fig1} (a) and (b), the structural 
heterogeneity has a large influence on the single neuron dynamics, at least at finite $K$ and
finite investigation times. In particular, 
in analogy with experiments \citep{gentet2010, mongillo2018} we considered
 a situation where the inhibitory drive prevails on the excitatory one. 
In this condition microscopically the neural populations splits in three groups: 
silent neurons, definitely sub-threshold; bulk neurons, which are fluctuation driven;
and mean driven outlier neurons. In particular,  excitatory (inhibitory) 
neurons with low (high) intra-population in-degrees are silenced due to the prevalence 
of synaptic inhibition. The silent neurons represent 6-10 \% of the whole population in agreement with experimental results 
for the mice cortex \citep{o2010}. Bulk neurons have in-degrees in proximity of the 
median and their firing rates approach the MF solution $\overline{R}_0^{(\alpha)}$ for increasing $K$.
Outlier neurons represent a minority group almost disconnected from their own population,
whose asymptotic behaviour for $K >> 1$ is controlled by the sign of the effective mean input current.

\vspace{0.5cm}
\subsection{Coherent Dynamics}
\vspace{0.5cm}

The emergence of COs is observable in this balanced network whenever the level of heterogeneity in the inhibitory population is
not too large, thus suggesting that the coherence among inhibitory neurons is fundamental to support collective rhythms
\citep{whittington2000}. Indeed we observed two main mechanisms leading to COs: one that can be identified as PING-like and another one 
as fluctuation driven. The PING-like mechanism is present whenever the excitatory neurons are able to deliver 
an almost synchronous excitatory volley that in turn elicits a delayed inhibitory one. 
The period of the COs is determined by the recovery time of the excitatory neurons from the stimulus received from the inhibitory population. This mechanism is characterized by a delay between the firing of the pyramidal cells and the interneuronal burst as reported also in many experiments \citep{buzsaki2012}. We have shown that this delay tends to vanish when the inhibitory action increases
leading the system from a balanced situation to a definitely sub-threshold condition where the neural activity
is completely controlled by fluctuations. In this latter case the excitatory and inhibitory neurons fire
almost simultaneously driven by the current fluctuations. These transform the relaxation dynamics towards a stable focus, observable
in the MF, to sustained COs via a mechanism previously reported for inhibitory networks \citep{matteo, bi2020}.
%An important point to stress is that COs in the considered balanced network emerge in absence of any synaptic or delay timescale
%analogously to what shown for purely inhibitory populations in \citep{matteo}.

The PING-like COs undergo period doubling cascades by varying $K$ and/or $I_0^{(e)}$ finally leading to collective chaos 
\citep{nakagawa1993, shibata1998}. The nature of this chaotic behaviour is definitely macroscopic
since it is captured by the neural mass model obtained within the MF formulation,
as shown by analysing the corresponding Lyapunov spectrum. This kind of chaos implies
irregular temporal fluctuations joined to a coherence at the spatial level over a large part of the network resembling 
coherent fluctuations observed across spatial scales in the neocortex \citep{volgushev2011,smith2008,okun2012,achermann2016}.
Collective (or coherent) chaos has been previously shown to be a ubiquitous feature
for balanced random spiking neural networks massively coupled, where $K$ is proportional to $N$ \citep{ullner2018, politi2018}.
Here, we have generalized such result to balanced random networks with sparse connectivity, where $K$ is independent by $N$.
Recently, it has been claimed that the presence of a structured feed forward connectivity in a random
network is needed to observe coherent chaos \citep{landau2018}. However, as evident from our results and those reported in  \citep{ullner2018, politi2018} coherent chaos can naturally emerge in a recurrent neural network in absence of any structured connectivity introduced {\it ad hoc} to promote collective behaviours. Furthermore, we have shown that collective chaos can emerge in random balanced networks with instantaneous synapses and in absence of any delay, see also \citep{ullner2018}.
%\blu{This seems to contrast with what happens in globally coupled neural systems, where collective chaos has been so far reported 
%in presence of some extrinsic time scale,  which can be a synaptic time scale, as in \citep{olmi2011} for two coupled 
%excitatory populations of LIF neurons and in  \citep{ceni2020} for two coupled 
%inhibitory QIF populations, or a time delay as shown in \citep{luccioli2010, pazo2016,devalle2018} for single inhibitory LIF and QIF %populations.}
 
Fluctuation driven COs are usually observable in our system as quasi-periodic collective motions 
characterized by two incommensurate frequencies. However, whenever the  current fluctuations become sufficiently strong the two frequencies can lock and give rise to a collective periodic motion. Furthermore, the locking region is characterized by a low level of synchrony in the network. These results resemble those reported in
\citep{meng2018} for two interconnected inhibitory neural networks subject to 
external uncorrelated noise. In particular, the authors have shown that uncorrelated noise sources
enhance synchronization and frequency locking  among the COs displayed by the two networks, despite the noise reduces the
synchrony among neurons within each network. At variance with \citep{meng2018}, in our case the noise sources
are intrinsic to the neural dynamics, but they can be as well considered as uncorrelated due to the sparseness in the
connections \citep{brunel1999,brunel2000}. Therefore we are reporting a new example of frequency locking among  
collective rhythms promoted by self-induced uncorrelated fluctuations. 
 
\red{
The frequencies of COs grows proportionally to the square root of the external excitatory DC current, as 
suggested by analytical arguments and confirmed by numerical simulations. This on one side allows,
simply by varying the parameters $I_0^{(e)}$ or $K$, to cover with our model a broad range of COs' frequencies 
analogous to those found experimentally in the cortex \citep{chen2017distinct}. On another side it
implies that the frequencies of COs diverge as $K^{1/4}$, while the average firing rates seem to converge to a common
value for sufficiently large $K$. These results seem to indicate that for large $K$ the network 
will become more and more unbalanced, with a prevalence of inhibition, while the amplitude of COs will tend to vanish.
However, this analysis is not conclusive and more detailed analysis are required to capture the asymptotic behaviour of the system in the limit $N >> K >> 1$.
}

\vspace{0.5cm}
\subsection{Future Developments}
\vspace{0.5cm}

\red{
The examined neural mass model has been derived by taking into account the random fluctuations due to the sparseness in the network connectivity only as a quenched disorder affecting the distribution of the effective synaptic couplings 
\citep{montbrio2015,matteo}. The current fluctuations can be correctly incorporated in a MF formulation
by developing a Fokker-Planck formalism for the problem, however this will give rise to a high (infinite) dimensional MF 
models \citep{brunel1999, brunel2000}. We are currently developing reduction formalisms for the Fokker-Planck equation 
to obtain low dimensional neural mass models which will include the intrinsic current fluctuations 
\citep{goldobin2021,noi}.}

%A further improvement would consist in including in a self-consistent manner the current fluctuations 
%in the MF formulation for balanced sparse networks. This has been done for homogeneous \citep{ullner2020}  and heterogeneous
%\citep{lerchner2006} LIF neurons. The self-consistent MF results have been obtained via refined iterative procedures
%based on the assumption that the synaptic input currents are the superposition of independent single neuron dynamics.
%However, such approaches neglect the correlations among the neurons and it is unclear
%how  they can be extended to system displaying COs.

\red{Relevant topics to investigate in the future to assess the generality of the reported results are 
their dependence on the chosen spiking neuron model and network architecture. 
In particular, for random networks it is important to understand the role played by
the distribution of the in-degrees, this also in view of the recent findings reported in
\citep{klinshov2021}.}

\section*{Author Contributions}
HB and MdV performed the simulations and data analysis. 
MdV and AT were responsible for the state-of-the-art review and the paper write-up.
All the authors conceived and planned the research.

\section*{Funding}
AT received financial support by the Excellence Initiative I-Site Paris Seine (Grant No ANR-16-IDEX-008) (together with HB), by the Labex MME-DII (Grant No ANR-11-LBX-0023-01) and by the ANR Project ERMUNDY (Grant No ANR-18-CE37-0014) (together with MdV), all part of the French programme ``Investissements d'Avenir''.

\section*{Acknowledgments}
The authors acknowledge extremely useful discussions with D.G. Goldobin, G. Mongillo,  E. Montbri\'o, S. Olmi, and A. Politi.
 
\section*{Conflict of Interest Statement}
The authors declare that the research was conducted in the absence of any commercial or financial relationships that could be construed as a potential conflict of interest.

%\section*{Supplemental Data}
% \href{http://home.frontiersin.org/about/author-guidelines#SupplementaryMaterial}{Supplementary Material} should be uploaded separately on submission, if there are Supplementary Figures, please include the caption in the same file as the figure. LaTeX Supplementary Material templates can be found in the Frontiers LaTeX folder.

\section*{Data Availability Statement}
The numerical programs and datasets for this study are availabl upon request.
% Please see the availability of data guidelines for more information, at https://www.frontiersin.org/about/author-guidelines#AvailabilityofData

%\bibliographystyle{frontiersinHLTH&FPHY}

\bibliographystyle{frontiersinSCNS_ENG_HUMS}

 % for Science, Engineering and Humanities and Social Sciences articles, for Humanities and Social Sciences articles please include page numbers in the in-text citations
 
\bibliography{lif_ost_cd}
%%% Make sure to upload the bib file along with the tex file and PDF
%%% Please see the test.bib file for some examples of references
 
%%% Please be aware that for original research articles we only permit a combined number of 15 figures and tables, one figure with multiple subfigures will count as only one figure.
%%% Use this if adding the figures directly in the mansucript, if so, please remember to also upload the files when submitting your article
%%% There is no need for adding the file termination, as long as you indicate where the file is saved. In the examples below the files (logo1.eps and logos.eps) are in the Frontiers LaTeX folder
%%% If using *.tif files convert them to .jpg or .png
%%%  NB logo1.eps is required in the path in order to correctly compile front page header %%%

%%% If you are submitting a figure with subfigures please combine these into one image file with part labels integrated.
%%% If you don't add the figures in the LaTeX files, please upload them when submitting the article.
%%% Frontiers will add the figures at the end of the provisional pdf automatically
%%% The use of LaTeX coding to draw Diagrams/Figures/Structures should be avoided. They should be external callouts including graphics.

\end{document}